\documentclass[reprint, superscriptaddress, amsmath, amssymb, aps, prc, noeprint]{revtex4-2}

\makeatletter
\renewcommand{\frontmatter@abstractwidth}{0.8\textwidth}
\makeatother

\usepackage{graphicx}
\usepackage{dcolumn}
\usepackage{bm}
\usepackage{orcidlink}
\usepackage{tabularx}

\usepackage{hyperref}

\begin{document}

\preprint{APS/123-QED}

\title{\textbf{Impact of neutron captures on the r-process: \\ I. Propagation of statistical model uncertainties}}

\author{Yu An\,\orcidlink{0009-0009-5919-6718}}
\email{yu.an@tu-darmstadt.de}
\affiliation{\mbox{Institut für Kernphysik, Technische Universität Darmstadt, Schlossgartenstr. 2, Darmstadt 64289, Germany}}
\affiliation{Helmholtz Forschungsakademie Hessen f\"ur FAIR (HFHF), Schlossgartenstra{\ss}e 2, Darmstadt 64289, Germany}

\author{Jan Kuske\,\orcidlink{0009-0005-5121-7343}}%
\affiliation{\mbox{Institut für Kernphysik, Technische Universität Darmstadt, Schlossgartenstr. 2, Darmstadt 64289, Germany}}

\author{Almudena Arcones\,\orcidlink{0000-0002-6995-3032}}
\affiliation{\mbox{Institut für Kernphysik, Technische Universität Darmstadt, Schlossgartenstr. 2, Darmstadt 64289, Germany}}
\affiliation{Helmholtz Forschungsakademie Hessen f\"ur FAIR (HFHF), Schlossgartenstra{\ss}e 2, Darmstadt 64289, Germany}
\affiliation{GSI Helmholtzzentrum für Schwerionenforschung GmbH, Planckstr. 1, Darmstadt 64291, Germany}
\affiliation{Max-Planck-Institut für Kernphysik, Saupfercheckweg 1, 69117 Heidelberg, Germany}

\date{\today}

\begin{abstract}
\setlength{\parindent}{0pt}
\textbf{Background:} The r-process is responsible for producing about half of the elements heavier than iron in the Universe, but the underlying nuclear physics is still largely inaccessible to experiments and relies on theoretical calculations.

\textbf{Purpose:} We aim to quantify the impact of statistical model uncertainties on the r-process by constructing a large set of neutron-capture rates within the Hauser-Feshbach framework with varied statistical model parameters. Propagating this set of neutron-capture rates to r-process calculations of various astrophysical sites allows us to link abundance features to nuclear uncertainties.

\textbf{Methods:} We calculate neutron-capture reactivities across the nuclear chart using different nuclear level density~(NLD) and gamma-ray strength function~(GSF) prescriptions. We study their impact on r-process calculations for individual representative trajectories, mass-integrated ejecta from a neutron-star-merger simulation, and a broad set of 10,000 parametric trajectories with a wide range of electron fractions and entropies.

\textbf{Results:} The final abundances are generally more sensitive to the NLDs than to the GSFs. Selected NLD models produce pronounced abundance changes between the lanthanides and the third peak. Although the variation is reduced in the mass-integrated total ejecta, a subpeak at $A\simeq172$--$178$ and systematic abundance changes at $A\simeq180$--$194$ persist. The parametric trajectories show that the NLD sensitivity is generally greater in the cold r-process regime, where neutron captures continue after photodissociations have become ineffective. Additional conditions with large NLD sensitivities are found at intermediate electron fractions, where small changes in the matter flow can lead to significant differences in the final abundances.

\textbf{Conclusions:} Systematic differences in neutron-capture rates between NLD prescriptions affect the final r-process abundances, particularly near nuclear shell closures. Therefore, improved experimental constraints and theoretical descriptions of NLDs in these regions are needed. The dependence of this sensitivity on astrophysical conditions highlights the need to assess nuclear uncertainties across a broad range of nucleosynthesis environments.

\end{abstract}

\maketitle

\section{Introduction}
\label{sec:intro}
Over the past few decades, our understanding of the rapid neutron-capture process (r-process), which is responsible for the production of roughly half of the elements heavier than iron, has advanced considerably \citep[see, e.g.,][]{Arnould2007_RprocessStellarNucleosynthesis, Horowitz:2018ndv, Cowan:2019pkx, Arcones:2022jer, 2026EPJA...62..105T}. Nevertheless, two major questions remain: the explosive astrophysical sites in which the r-process occurs and the nuclear physics of the neutron-rich nuclei involved. Neutron star mergers~(NSMs), which provide extremely neutron-rich conditions, have been observationally confirmed as an r-process site through the multi-messenger observation of the kilonova AT2017gfo \citep{Abbott2017_MultimessengerObservationsBinary, Drout2017_LightCurvesNeutron, Kilpatrick2017_ElectromagneticEvidenceThat}. However, Galactic chemical evolution models suggest that additional prompt r-process sources are needed to explain the enrichment of the early Galaxy \citep[see, e.g.,][]{Argast2004_NeutronStarMergers, 2014MNRAS.438.2177M, Cote2019_NeutronStarMergers, Molero2023_OriginNeutronCapture}. Suggested additional sites include magneto-rotational core-collapse supernovae \citep[see, e.g.,][]{Winteler2012_MagnetorotationallyDrivenSupernovae, Reichert2021_NucleosynthesisMagnetorotationalSupernovae}, collapsars \citep[see, e.g.,][]{Siegel2019_CollapsarsMajorSource}, and magnetar giant flares \citep[see, e.g.,][]{Patel2025_DirectEvidenceRprocess}. 

Nuclear physics inputs constitute another major source of uncertainty in r-process calculations, in particular nuclear masses \citep[see, e.g.,][]{Kratz1993_IsotopicRProcessAbundances, Goriely2001_NuclearMassesPProcesses, Wanajo:2004ai, Arcones:2010dz, Mumpower:2015hva, Martin:2015xql, deJesusMendoza-Temis:2014owk, Vassh2021_MarkovChainMonte, Giuliani:2024ppp, Kuske:2025tsm, Martinet2025_ImpactMassUncertainties}, beta decay \citep[see, e.g.,][]{Marketin2016_LargescaleEvaluation$ensuremathbeta$decay, Lund2023_InfluenceBetaDecay}, neutron capture \citep[see, e.g.,][]{Goriely1997_DirectNeutronCaptures, Surman1997_SourceRareEarthElement, Goriely:1998utv, Mumpower2012_InfluenceNeutronCapture, Surman2014_SensitivityStudiesWeak, Nikas:2020zcc, Wiedeking:2024bty}, and fission properties \citep[see, e.g.,][]{Goriely2013_NewFissionFragment, Eichler2015_RoleFissionNeutron, Vassh2019_UsingExcitationenergyDependent, Giuliani2020_Fission$r$processNucleosynthesis, Holmbeck2023_SuperheavyElementsKilonovae}.  Experiments for nuclear masses and beta decays have greatly improved in recent years and are reaching the r-process path at $N=50$ and $N=82$ \citep[see, e.g.,][]{Xu:2014hla, Atanasov:2015eza, Lorusso:2015cna, Welker:2017eja}. Also neutron-capture rates towards the r-process are getting constrained by experiments, however until now only through indirect methods. The Oslo and beta-Oslo methods \citep[see, e.g.,][]{Schiller2000_ExtractionLevelDensity, Spyrou2014_NovelTechniqueConstraining, Larsen2019_NovelTechniquesConstraining, Greaves2025_ProbingLimitsStatistical} can extract nuclear level densities (NLDs) and gamma-ray strength functions (GSFs), which are subsequently used to constrain statistical neutron-capture rates within the Hauser-Feshbach framework~\citep{Hauser:1952zz}. Surrogate-reaction approaches, such as $(d,p\gamma)$ and $(p,p'\gamma)$ measurements~\cite{Escher:2018zhz, Ratkiewicz:2018wpx, Zhang:2025jta}, instead populate the same compound nucleus through an alternative entrance channel and constrain its $\gamma$-decay probability, thereby providing information on the compound contribution to the desired $(n,\gamma)$ cross section. While the Trojan horse \citep{Baur1986_BreakupReactionsIndirect} and asymptotic normalization coefficient \citep{Xu1994_OverallNormalizationAstrophysical} methods can provide information relevant to direct capture, and Coulomb dissociation can constrain the inverse photodissociation reaction \citep{Baur1986_CoulombDissociationSource}. Proposed concepts for neutron targets may allow for direct measurements of neutron-capture rates in the near future \citep[see, e.g.,][]{Reifarth2014_MeasurementsNeutroninducedReactions, Reifarth2017_SpallationbasedNeutronTarget, Tarifeno-Saldivia2026_DirectNeutronReactions}.

Current r-process nucleosynthesis calculations still rely largely on theoretical neutron-capture rates. These are commonly based on the statistical Hauser-Feshbach model and require several nuclear-physics inputs, most importantly nuclear masses, NLDs, and GSFs. In addition, the optical model potential (OMP) \citep[see, e.g.,][]{Kedia:2026usu} and direct captures \citep[][]{Goriely1997_DirectNeutronCaptures, Sieja2021_ShellmodelBasedStudy} can have an impact on r-process abundances, although comparatively minor. Because the impact of nuclear masses on neutron-capture rates and r-process abundances has already been extensively studied, in this work we focus on the uncertainties arising from different NLD and GSF models.

The main impact of neutron captures occurs during the decay towards stability after $(n,\gamma)-(\gamma,n)$ equilibrium. This effect is most pronounced in astrophysical conditions with a fast temperature decline, in which the decay phase starts before the majority of neutrons are depleted. In such a cold r-process scenario, the final abundances are governed by a competition between beta decays and neutron captures \citep[see, e.g.,][]{Meyer:1994gn, Surman1997_SourceRareEarthElement, Arcones:2010dz, Nikas:2020zcc, Li:2026sur}. Recently, \citet{Xu2026_ImpactNuclearLevel} have investigated the impact of NLDs on the rare earth region. They found that NLD differences in even-$A$ nuclei redirect the reaction flow, accelerating the rare-earth peak formation. We extend this investigation beyond the rare-earth region and consider a wider set of astrophysical conditions, including a full mass-integrated neutron star merger ejecta. Our analysis treats correlations between nuclei consistently by applying different NLD prescriptions to all nuclei at the same time. This approach prevents unphysical assumptions for neighboring nuclei, but the resulting parameter ranges could be artificially limited if the underlying models have systematic shortcomings. An alternative approach is to assume uncorrelated uncertainties, e.g., by varying all reaction rates independently from each other within their experimental or theoretical uncertainties in a Monte Carlo framework \citep[see, e.g.,][]{Mumpower:2015hva, Bliss2020_NuclearPhysicsUncertainties, Psaltis2022_ConstrainingNucleosynthesisNeutrinodriven}. 

This manuscript is structured as follows: In Section~\ref{sec:method}, we briefly describe the nucleosynthesis and reaction rate calculations. Section~\ref{sec:rprocess_ng} provides a general overview on the effect of neutron captures on the r-process. In Section~\ref{sec:results}, we investigate the impact of different NLDs and GSFs under a variety of astrophysical trajectories from different representative r-process sites, the mass-integrated ejecta of a neutron star merger (NSM) simulation, and a broad parametric study. In Section~\ref{sec:conclusions}, we summarize our main findings.

\section{Method}
\label{sec:method}

\subsection{Nucleosynthesis network and astrophysical conditions}
\label{sec:network}

Nucleosynthesis calculations are carried out with the open-source nuclear reaction network WinNet~\cite{Reichert:2023xqy} including the nuclear physics input as in \citet{Kuske:2025ffp}. Additional neutron-capture reactivities $\langle \sigma v \rangle$, are calculated with the nuclear reaction code TALYS-2.0~\cite{Koning:2023ixl} within the Hauser-Feshbach statistical framework. The corresponding reverse rates are derived through detailed balance. The energy released by nuclear reactions, which feeds back into the temperature evolution, is taken into account through the entropy variation as described in \citet{Reichert:2023xqy}. Our nucleosynthesis calculations start at 7~GK with abundance distributions given by nuclear statistical equilibrium.

In order to understand the general impact of neutron captures on the r-process, we investigate a variety of astrophysical conditions, including representative trajectories from different r-process sites, the total ejecta of a NSM simulation, and a broad parametric study. The representative trajectories include a NSM dynamical ejecta trajectory and a post-NSM accretion disk (NSM-DISK) trajectory. For the individual and integrated NSM trajectories, we use the LS255 model from \citet{Jacobi:2023olu} and \citet{Ricigliano:2024lwf}, which simulates an equal-mass binary merger, forming a relatively stable massive neutron star that survives throughout the simulation time. The NSM-DISK trajectory is adopted from the post-NSM accretion disk simulation by \citet{Wu:2016pnw}. The parametric model follows the method described in \citet{Kuske:2025ffp}. Starting with an initial electron fraction $Y_{e,0}$ and entropy $s_0$, the nucleosynthesis follows a density parametrization that depends on an expansion timescale $\tau$. For all parametric models considered in this work, the initial electron fraction $Y_\mathrm{e,0}$ is varied from 0.005 to 0.5 using 100 linearly spaced values, while the initial entropy $s_0$ is varied from 1 to 290~$k\mathrm{_B/nuc}$ using 100 logarithmically spaced values. Here, the expansion timescale is fixed to a typical value of $\tau = 4\,\mathrm{ms}$. These parameter combinations cover many of the conditions found in current simulations of r-process sites, while some parameter combinations also extend beyond the conditions currently predicted by simulations, cf. Fig.~9 in \citet{Kuske:2025ffp}.

\subsection{Statistical model}
\label{sec:statmodel}

Neutron-capture reactivities are calculated with TALYS-2.0~\cite{Koning:2023ixl} within the Hauser-Feshbach statistical framework~\cite{Hauser:1952zz} and based on different nuclear physics input for nuclear masses, OMPs, NLDs, and GSFs. The impact of the mass model on the r-process has already been extensively explored \citep[see, e.g.,][]{Kratz1993_IsotopicRProcessAbundances, Goriely2001_NuclearMassesPProcesses, Wanajo:2004ai, Arcones:2010dz, Mumpower:2015hva, Martin:2015xql, deJesusMendoza-Temis:2014owk, Vassh2021_MarkovChainMonte, Giuliani:2024ppp, Kuske:2025tsm, Martinet2025_ImpactMassUncertainties}, therefore we do not vary it here and  use FRDM2012~\cite{Moller:2015fba}, as in~\citet{Kuske:2025ffp}. Moreover, we found a very small impact of the OMP and thus fix it to the global Koning-Delaroche potential (see also \cite{Nikas:2020zcc} and \cite{Kedia:2026usu}). Therefore, we focus in this work on the impact of NLDs and GSFs as the dominant sources of statistical-model uncertainty.

\begin{figure}[!t]
    \centering
    \includegraphics[width=\columnwidth]{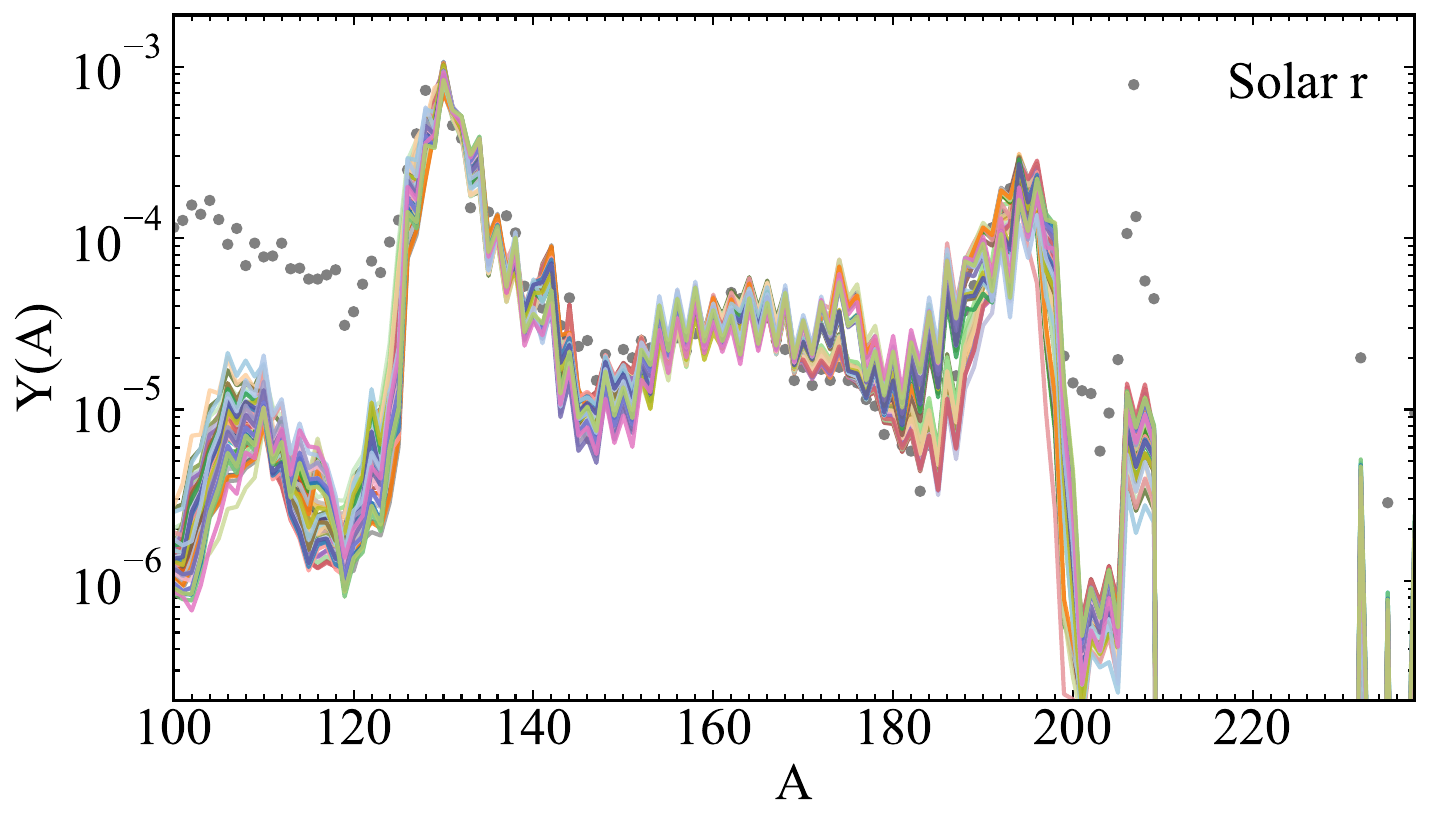}
    \caption{
    Final abundances based on the NSM trajectory~\cite{Jacobi:2023olu} as a function of mass number for all 60 combinations of the six NLD models (Table~\ref{tab:ld}) and ten GSF models (Table~\ref{tab:sf}). Gray dots show the solar r-process residual abundances~\citep{2008ARA&A..46..241S}, scaled to $Y(A=151)=2\times10^{-5}$.
    }
    \label{fig:nsm60comb}
\end{figure}

We consider all NLD and GSF models available in TALYS-2.0 (see Tables~\ref{tab:ld} and \ref{tab:sf}), resulting in a total of 60 different statistical model combinations. For each combination we have calculated the neutron-capture reactivities for all nuclei (from $Z=8$ to $118$). Figure~\ref{fig:nsm60comb} shows the resulting spread in the final abundances based on the NSM trajectory. 

In the following analysis, we adopt LD5+SF8 as the reference model combination, motivated by its favorable performance in previous systematic studies~\cite{Goriely:2008zz, Goriely:2018gig} and its status as the most trusted combination in the TENDL evaluation~\cite{Rochman:2024yua}. Although some of the NLD and GSF models considered here show poor agreement with experimental data \citep[see, e.g.,][]{Wiedeking:2024bty, Goriely:2019upm, Goriely:2022flp, Markova:2025iwu}, their predictive performance for neutron-rich nuclei relevant to the r-process remains uncertain. Excluding them based solely on their agreement near stability could therefore artificially narrow the estimated model spread in the region that matters most. We consequently use the full set of models to obtain a conservative estimate of the statistical-model uncertainty, without assuming that all models (and even all combinations of NLD and GSF) are equally reliable.

\begin{table}[t]
    \caption{Nuclear level density (NLD) models used in this work. The reference model is marked in bold.}
    \label{tab:ld}
    \centering
    \begin{tabularx}{\columnwidth}{lX}
    \hline
    \hline
    ID & Model \\
    \hline
    LD1 & Constant-temperature plus Fermi gas model (Cst-T)~\cite{Koning:2008ygi} \\
    LD2 & Back-shifted Fermi gas model (BSFG)~\cite{Demetriou:2001cvr} \\
    LD3 & Generalized Superfluid model (GSM)~\cite{osti_5770504, 1993PhRvC..47.1504I} \\
    LD4 & Skyrme-HFB from numerical tables~\cite{Goriely:2001zz} \\
    \textbf{LD5} & \textbf{Skyrme-HFB combinatorial from numerical tables (HFB+comb.)}~\cite{Goriely:2008zz} \\
    LD6 & Temperature-dependent Gogny-HFB combinatorial from numerical tables (THFB+comb.)~\cite{Hilaire:2012tb} \\
    \hline
    \hline
    \end{tabularx}
\end{table}

\begin{table}[t]
    \caption{Same as Table~\ref{tab:ld} for the $\gamma$-ray strength function (GSF) models.}
    \label{tab:sf}
    \centering
    \begin{tabularx}{\columnwidth}{lX}
    \hline
    \hline
    ID & Model \\
    \hline
    SF1 & Kopecky-Uhl generalized Lorentzian (GLO)~\cite{Kopecky:1990zz} \\
    SF2 & Brink-Axel Lorentzian (SLO)~\cite{1957NucPh...4..215B, Axel:1962zz} \\
    SF3 & Hartree-Fock BCS tables (HF-BCS)~\cite{Goriely:2002cx} \\
    SF4 & Skyrme-HFB with QRPA (HFB+QRPA)~\cite{Goriely:2003rz} \\
    SF5 & Hybrid model (GH)~\cite{Goriely:1998utv} \\
    SF6 & Temperature-dependent Skyrme-HFB with QRPA (THFB+QRPA)~\cite{Goriely:2003rz} \\
    SF7 & Temperature-dependent RMF (TRMF)~\cite{Daoutidis:2012iz} \\
    \textbf{SF8} & \textbf{Gogny-HFB plus QRPA with D1M interaction (D1M+QRPA)}~\cite{Goriely:2018gig} \\
    SF9 & Simplified Modified Lorentzian Model (SMLO)~\cite{Goriely:2019upm} \\
    SF10 & Skyrme-HFB plus QRPA with BSk27 interaction (BSk27+QRPA)~\cite{Xu:2021xet} \\
    \hline
    \hline
    \end{tabularx}
\end{table}

\section{When are neutron captures relevant during the r-process?}
\label{sec:rprocess_ng}
\begin{figure*}[!t]
    \centering
    \includegraphics[width=1.0\textwidth]{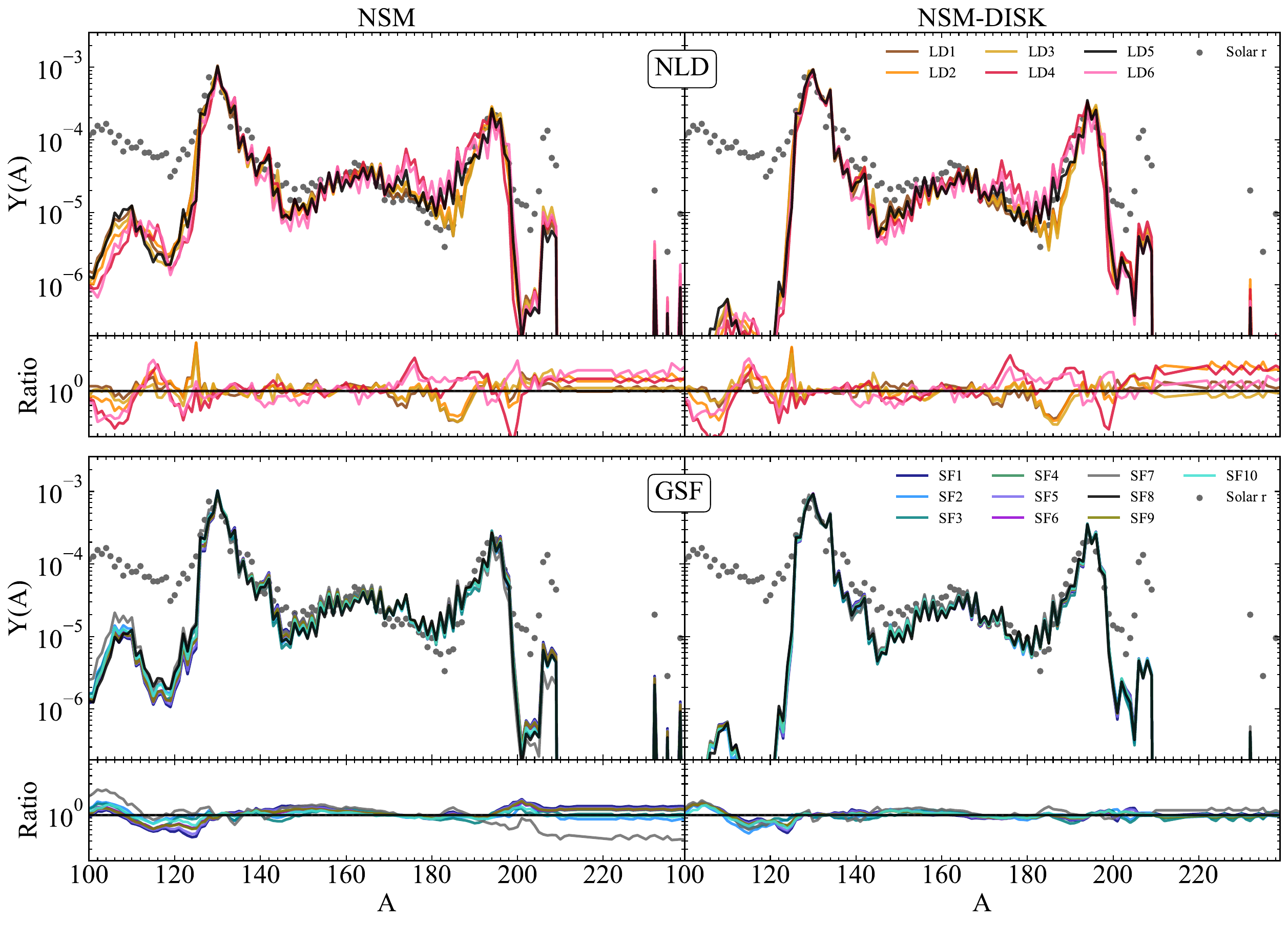}
    \caption{Final abundances at 1\,Gyr based on neutron-capture rates with SF8 and different NLDs  (upper row) and with LD5 and different GSFs (lower row). Left and right columns correspond to two different astrophysical trajectories from dynamical NSM~\cite{Jacobi:2023olu} and NSM disk ejecta~\cite{Wu:2016pnw}, respectively. The abundance ratios are relative to the reference models (LD5 and SF8). Solar r-process residual abundances~\cite{2008ARA&A..46..241S} are normalized to $Y(A=151)=2\times10^{-5}$. 
    } 
    \label{fig:overview}
\end{figure*}
Before discussing our results, we summarize here the role of neutron captures  during the different phases of r-process. Our calculations start from nuclear statistical equilibrium at $T=7\,\mathrm{GK}$. Once the temperature drops to about $3\,\mathrm{GK}$, charged-particle reactions become too slow to overcome the Coulomb barrier and freeze out, ending this phase and starting the r-process. The  high neutron densities found in r-process conditions, however, keep neutron captures and photodissociations in equilibrium. During this (n,$\gamma$)-($\gamma$,n) equilibrium, the abundance ratio of neighboring isotopes in an isotopic chain is given by \citep[see, e.g.,][]{Cowan:2019pkx}:
\begin{align}
\frac{Y(Z,A+1)}{Y(Z,A)} = &\left(\frac{2\pi\hbar^2}{m_\mathrm{u} k_\mathrm{B} T}\,\frac{A+1}{A}\right)^{3/2}\frac{G_{Z,A+1}(T)}{2\,G_{Z,A}(T)}\notag\\
&\exp\!\left(\frac{S_\mathrm{n}(Z,A+1)}{k_\mathrm{B} T}\right) N_\mathrm{n}\,,
\label{eq:equilibrium}
\end{align}
where $G_{Z,A}(T)$ is the temperature-dependent partition function of nucleus $(Z,A)$, $S_\mathrm{n}$ is the one-neutron separation energy, and $N_\mathrm{n}:=Y_\mathrm{n}\rho N_A$ is the neutron number density. Equation~\eqref{eq:equilibrium} depends on $N_\mathrm{n}$, $T$, and, exponentially, on $S_n$, but not on the neutron-capture cross section itself. However, the nuclear level density  already plays a role at this stage, since it enters through the partition functions $G_{Z,A}(T)$ and thereby affects the equilibrium abundances.
As temperature decreases, two  points become relevant to understand the late r-process evolution and the role of neutron captures. The r-process freeze-out (RFO) is reached once the available neutrons are exhausted, $Y_\mathrm{n}/Y_\mathrm{seed}=1$, and further neutron captures become small or negligible. Additionally, the $(n,\gamma)$-$(\gamma,n)$ equilibrium itself ends once photodissociation can no longer keep pace with neutron capture, $\tau_{(\gamma,n)}>\tau_{(n,\gamma)}$, where $\tau$ denotes the average reaction timescale \citep[cf.,][]{Arcones:2010dz}. We refer to this moment as the equilibrium freeze-out (EFO).

Whether EFO occurs before or after RFO determines the character of the r-process~\cite{Arcones:2010dz}. If EFO occurs after RFO (hot r-process), the (n,$\gamma$)-($\gamma$,n) equilibrium survives until the neutrons  run out. The freeze-out abundances are then still well described by Eq.~\eqref{eq:equilibrium}, and the  neutron-capture rate has small impact, since essentially no neutrons remain once the system drops out of equilibrium. If instead EFO occurs before RFO (cold r-process), the equilibrium breaks down while still enough neutrons are present. Neutron captures then keep competing with beta decays and beta-delayed neutron emissions while matter decays to stability. The resulting abundances depend directly on the  neutron captures. It is precisely in this post-equilibrium, pre-freeze-out window (reached by a large fraction of realistic r-process conditions) that the choice of NLD and GSF models directly shapes the final abundances discussed in Sec.~\ref{sec:results}.

\section{Results}
\label{sec:results}
\subsection{Overview of statistical model impact on the r-process}
We have explored the impact of varying NLDs and GSFs on the r-process abundances in Fig.~\ref{fig:overview} for two representative trajectories (dynamical NSM ejecta and NSM-DISK outflow) and for six NLD prescriptions (upper row, using SF8) and ten GSF prescriptions (lower row, using LD5). The lower panels show the relative variations compared to the reference model (LD5+SF8). Generally, the different NLDs result in larger abundance variations than the GSF variations.
The NLD variations result in about one order of magnitude difference below the second peak ($A\lesssim130$). There are also significant differences in the region between the rare-earth and third peak, most notably around $A\sim172-178$ \citep[see also][]{Xu2026_ImpactNuclearLevel}. In both trajectories, the third r-process peak remains relatively robust in position, overall shape and height, while the left side at $A\sim180-194$ changes significantly. Additionally, the Pb peak at $A\sim206$-$208$ and the actinide peak at $A\sim232$-$238$ also exhibit comparatively limited sensitivity to the NLD prescription. In contrast, most GSF prescriptions produce very similar abundance patterns, differing less than a factor of 2. Larger deviations are found for SF7, particularly in the heavy-mass region. The distinctive behavior of SF7 may be associated with its enhanced $(n,\gamma)$ reactivities near shell closures and reduced reactivities between them, which suppress the matter flow toward heavier nuclei.

Among the statistical-model ingredients considered here, the NLD is therefore the dominant source of uncertainty, with a sensitivity that strongly depends on mass number. The major r-process peaks remain comparatively robust to this choice, whereas the regions between the peaks are most affected. In addition, we notice that in agreement with \citet{Xu2026_ImpactNuclearLevel}, the three phenomenological NLDs (LD1-LD3) result in very similar abundances, whereas the microscopical NLDs (LD4-LD6) differ from one another more significantly.
    
\subsection{Impact of NLDs on the r-process}
\label{sec:impactOfNLDs}

\begin{figure}[!t]
    \includegraphics[width=\columnwidth]{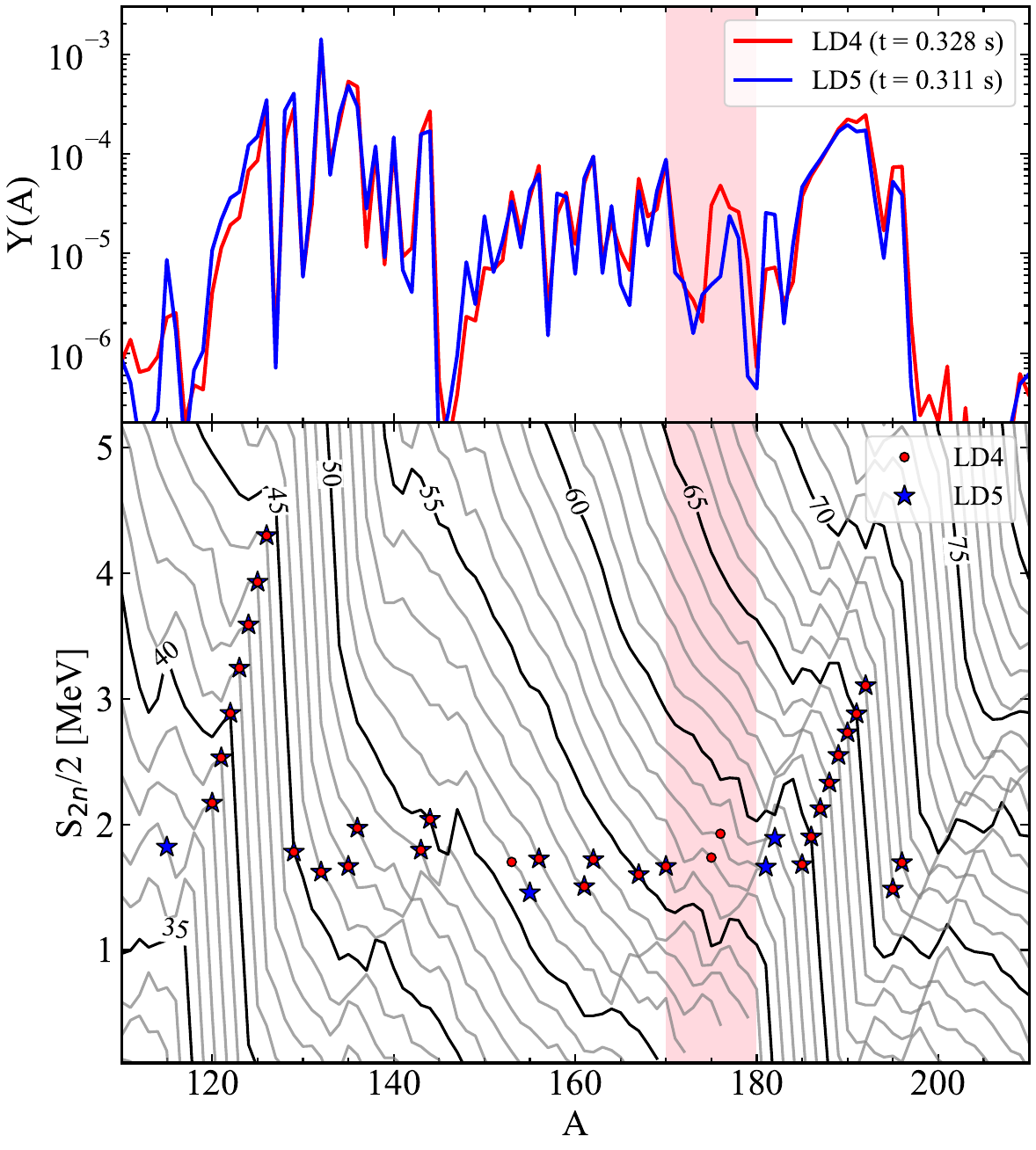}
    \caption{
    RFO abundances~(upper panel) and two-neutron separation energies of the most abundant isotopes of each isotopic chain~(lower panel) as a function of mass number for LD4 (red) and LD5 (blue). Black lines correspond to proton numbers from $Z=30$ on the left and increase in steps of 5, while the other isotopic chains are shown as gray lines. The RFO times are given in the upper legend. The pink band indicates the region where the $A\sim172-178$ subpeak forms.}
    \label{fig:path_freezeout}
\end{figure}
We now focus on the impact of NLDs and examine the origin of the pronounced abundance features between $A=172-178$ and $180-194$ for the  representative NSM trajectory (upper left panel in Fig.~\ref{fig:overview}), as well as for the mass-integrated total ejecta of all trajectories of the NSM simulation \citep{Jacobi:2023olu}.

The $A\sim172-178$ subpeak has its maximum near $A\simeq174$ for LD4 and LD6, where these NLD prescriptions enhance the final abundances by more than a factor of three relative to the reference model (LD5). The subpeak forms during the time window between EFO and RFO, where there is a competition between neutron captures and beta decays (see Sec.~\ref{sec:rprocess_ng}). Throughout this post-EFO window, LD4 predicts systematically lower $(n,\gamma)$ reactivities than LD5, which slows the depletion of the neutron reservoir and delays RFO. During this additional time, slower neutron captures favor material accumulating below the third r-process peak in LD4, while beta decays afterwards shift the r-process path slightly closer to stability before RFO is reached. Consequently, the RFO abundances (upper panel in Fig.~\ref{fig:path_freezeout}) differ between the two NLD models, in particular around $A\sim170-180$, where the neutron separation energies ($S_\mathrm{2n}/2$) show a saddle-shaped feature (pink band in lower panel of Fig.~\ref{fig:path_freezeout}). This leads to the r-process path in the La and Ce isotopic chains to be localized at smaller mass numbers (red dots) around $A\sim170-180$ in LD4, providing the precursor of the final subpeak.
\begin{figure*}[!t]
    \centering
    \includegraphics[width=\textwidth]{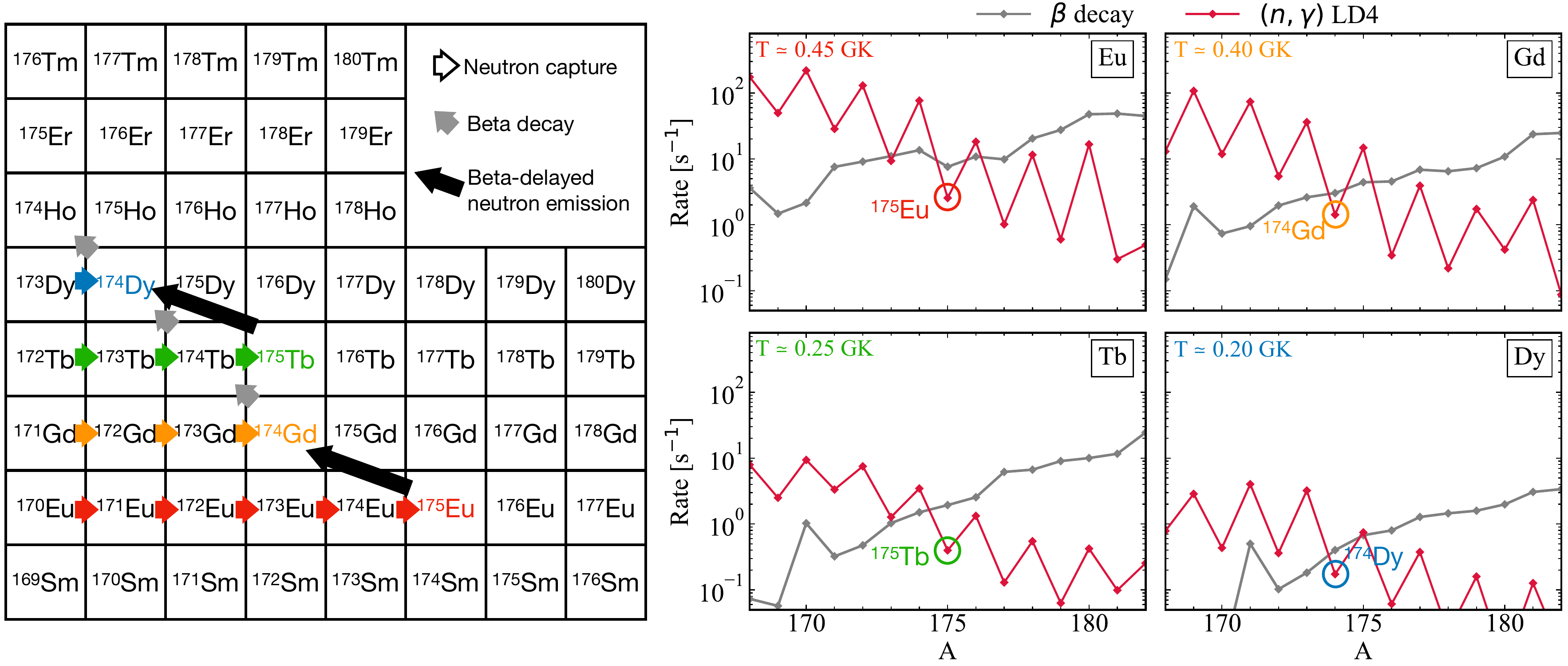}
    \caption{
    Left panel: Schematic reaction flows associated with the formation of the $A=174$ subpeak for the NSM trajectory with LD4. Right panels: beta decay and neutron-capture rates along the Eu, Gd, Tb, and Dy isotopic chains ($Z=63-66$), evaluated at the indicated temperatures, when each chain is most abundant relative to neighboring elements in the region. The most abundant isotopes of each isotopic chain (highlighted in both panels) are typically the lightest ones, for which the neutron-captures rates are slower than beta decays.
    }
    \label{fig:local_peak_flow}
\end{figure*}
During the post-freeze-out evolution, residual neutron captures, beta decays, and beta-delayed neutron emission further reshape the abundances, as schematically shown in the left panel of Fig.~\ref{fig:local_peak_flow}. The right panel shows the corresponding reaction rates ~$r=N_\mathrm{n}\langle \sigma v\rangle$ for beta decays and neutron captures for LD4 for the involved isotopic chains (Eu, Gd, Tb, and Dy), at the temperatures at which they are approximately most abundant relative to the neighboring elements. At $T=0.45\,\mathrm{GK}$, when the subpeak feature starts to form, $^{171}$Eu, $^{173}$Eu, and $^{175}$Eu are strongly populated. Neutron captures dominate over beta decays along the Eu chain up to $^{175}$Eu, where beta decays become faster than neutron captures. Beta-delayed neutron emission from $^{175}$Eu, together with neutron captures on lighter Gd isotopes, populate $^{174}$Gd as the temperature approaches 0.4~GK. Subsequent beta decays of $^{174}$Gd feed $^{174}$Tb, followed by neutron captures to form $^{175}$Tb. Around $T\simeq0.25\,\mathrm{GK}$, beta decays dominates over neutron captures for $^{175}$Tb, and its beta-delayed neutron emission feeds $^{174}$Dy. By $T\simeq0.20\,\mathrm{GK}$, neutron captures on $^{174}$Dy are also slower than beta decays, limiting further mass growth and helping to preserve the accumulation at $A=174$ as the material decays toward stability. In the LD5 scenario, less material is present in this mass region at freeze-out. Although LD5 has larger intrinsic neutron-capture reactivities, its faster neutron consumption results in a lower late-time neutron density~$N_\mathrm{n}$ and therefore smaller effective capture rates. The same concentration of reaction flow does not develop, resulting in a less pronounced subpeak.

A similar enhancement is also observed in the NSM-DISK trajectory, whose initial electron fraction and entropy are comparable to those of the NSM trajectory. This result indicates that the localized NLD sensitivity in the $A\sim172-178$ region is not restricted to a single astrophysical trajectory. 

A broad NLD-dependent variation is also found for the mass interval $A\sim180-194$, i.e., on the left side of the third r-process peak. Relative to the reference model (LD5), LD4 and LD6 produce an enhancement in this region, whereas LD3 gives the strongest depletion. To understand this behavior, we compare LD3, LD4, and LD5 for the NSM trajectory. The differences develop mainly during the post-freeze-out evolution, when the temperature decreases to approximately $0.45\,\mathrm{GK}$. LD3 predicts generally larger $(n,\gamma)$ rates than LD5 for many nuclei near the r-process path in the $N\sim126$ region. After freezeout, the stronger residual neutron captures transfer more material towards higher mass numbers and thereby deplete the $A\sim180-194$ region. In contrast, LD4 predicts generally lower capture rates than LD5 in the $N\sim126$ region. This reduces the late-time flow towards the third peak and leaves more material in the $A\sim180-194$ region. Subsequent beta decays preserve the mass number, whereas beta-delayed neutron emission can shift part of the material toward slightly lower masses. 

A similar model dependence is also found in the representative NSM-DISK trajectory, suggesting that this behavior is not restricted to a single astrophysical condition.

\begin{figure}[!t]
    \centering
    \includegraphics[width=\columnwidth]{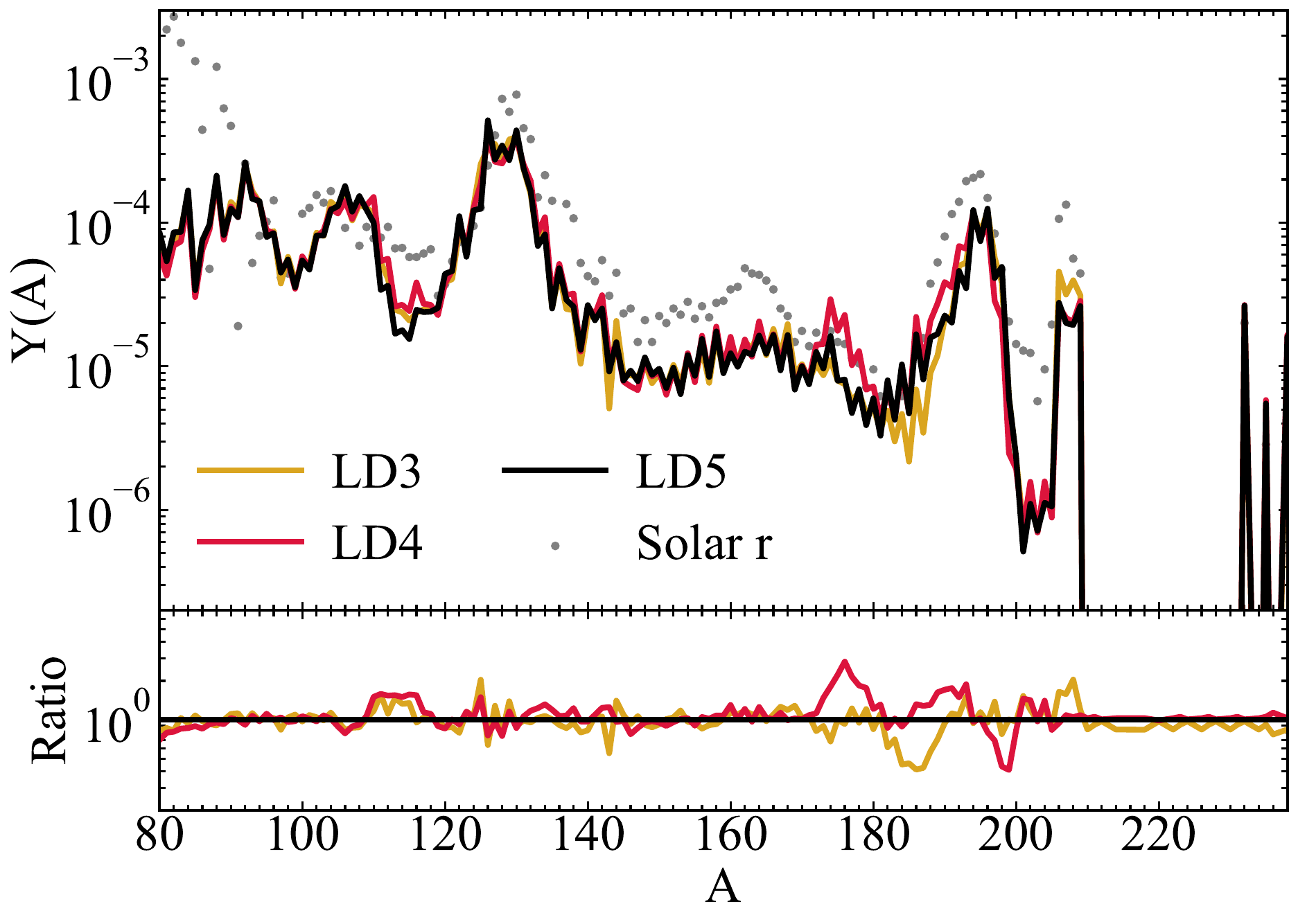}
    \caption{
    Final abundances at 1 Gyr, mass-integrated over all trajectories for the lower-extreme combination (LD3+SF8), upper-extreme combination (LD4+SF8), and reference combination (LD5+SF8) from the NSM simulation of \cite{Jacobi:2023olu}.
    }
    \label{fig:nsm_ld}
\end{figure}
\begin{figure*}[!t]
    \centering
    \includegraphics[width=\textwidth]{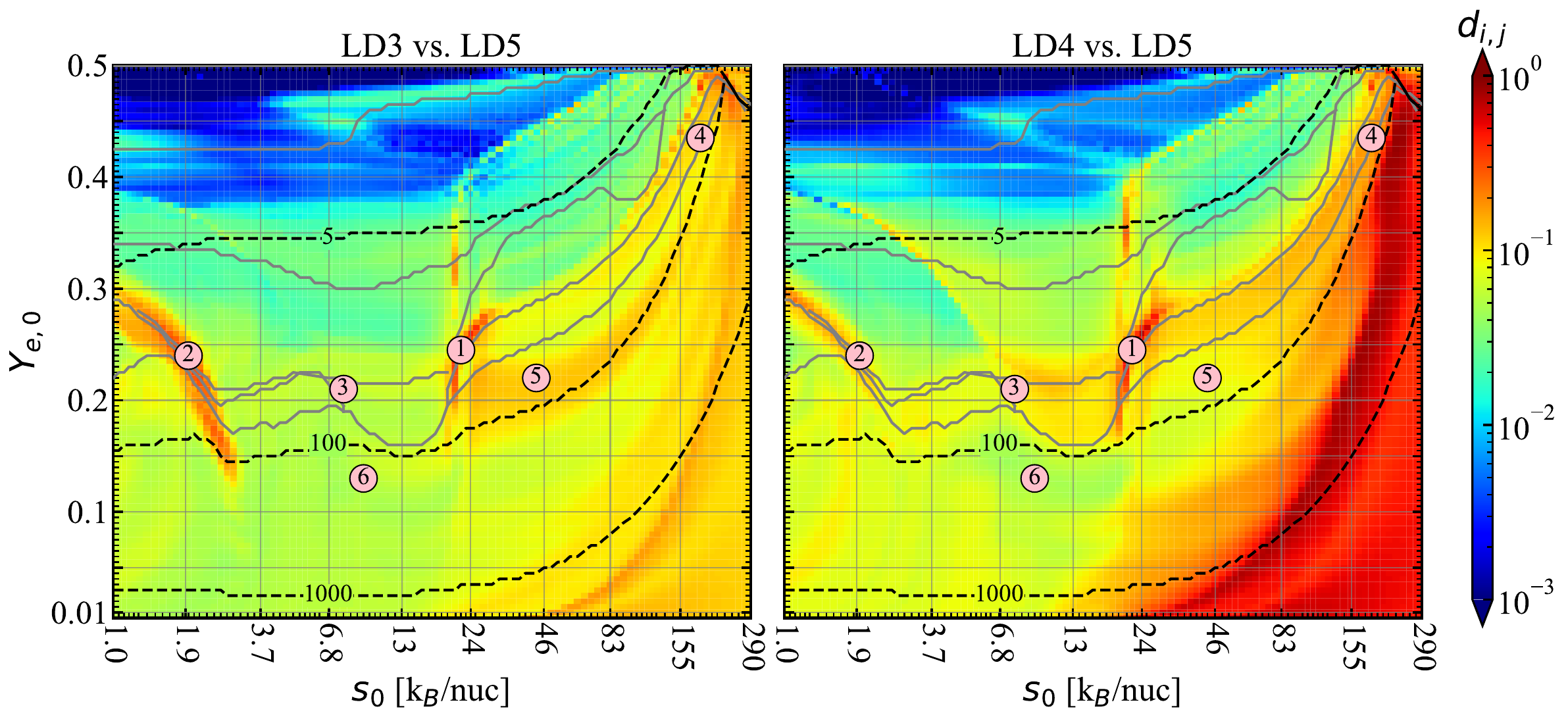}
    \caption{
    Pairwise abundance difference $d_{i,j}$ between LD3+SF8 and the reference LD5+SF8~(left), and between LD4+SF8 and the reference LD5+SF8~(right) across the ($Y_\mathrm{e,0}$, s$_0$) parameter space for an expansion timescale of $\tau$ = 4~ms. The gray solid lines and black dashed lines display the borders of the nucleosynthesis groups from \citet{Kuske:2025ffp} and the neutron-to-seed ratios at $T=3$~GK, respectively. The final abundances of six selected conditions (numbered points) are given in Fig.~\ref{fig:parametric_interested_points}.
    }
    \label{fig:parametric_ld}
\end{figure*}

\begin{figure*}
    \centering
    \includegraphics[width=0.9\textwidth]{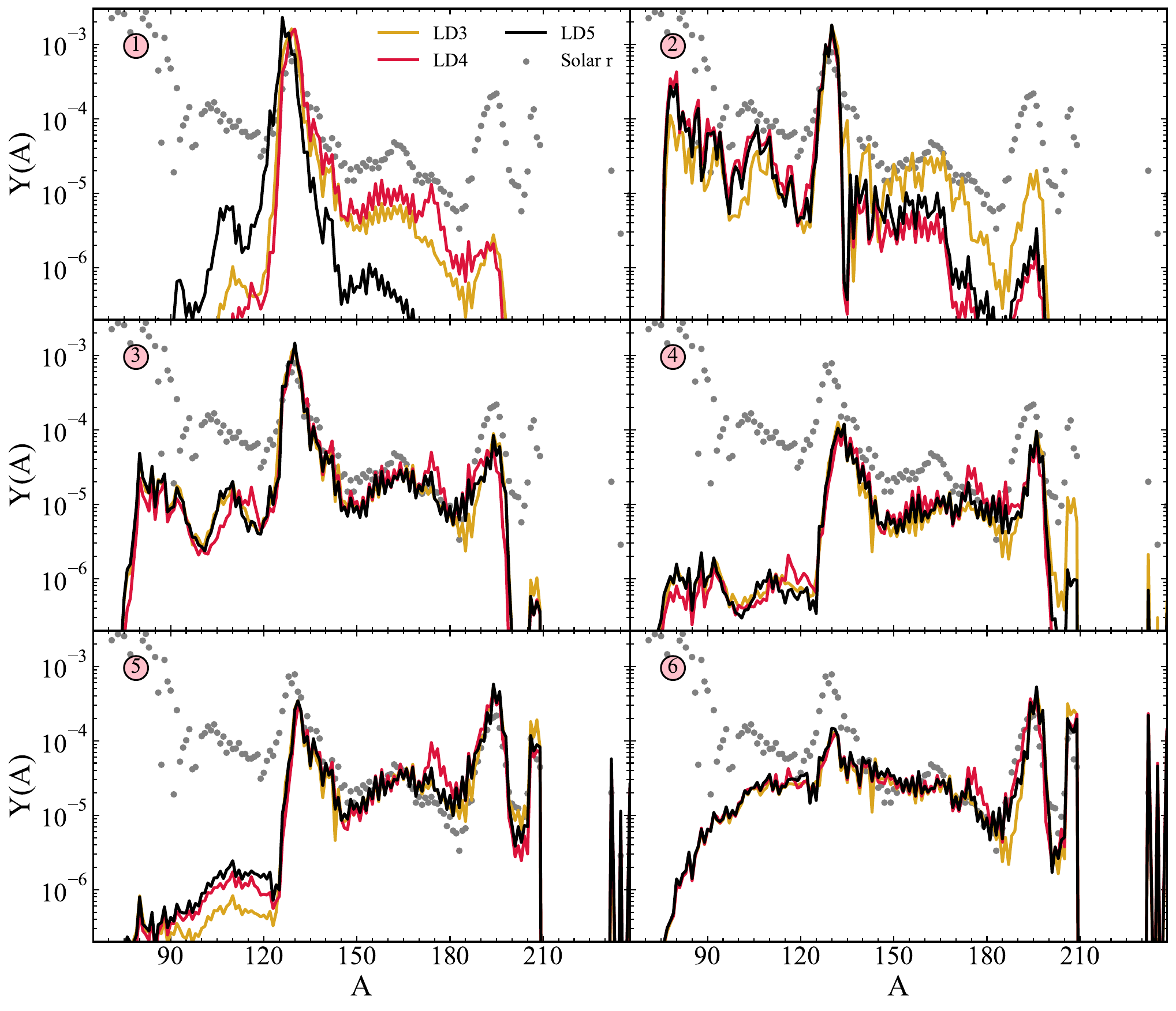}
    \caption{
    Final abundances for selected parameter conditions highlighted in Fig.~\ref{fig:parametric_ld}. 
    }
    \label{fig:parametric_interested_points}
\end{figure*}
The impact of the NLD prescription is further examined using the trajectory-integrated abundances, obtained by summing all (over 11,000) NSM trajectories from \citet{Jacobi:2023olu} (model LS255), weighted with their corresponding ejecta-masses in Fig.~\ref{fig:nsm_ld}. Compared with the representative trajectory, the integration smooths some of the pronounced variations found in individual trajectories. Nevertheless, a clear model dependence remains, particularly in the $A\sim172-178$ and $A\sim180-194$ regions. These mass regions are produced mainly by the neutron-rich dynamical ejecta, which span a broad $Y_\mathrm{e,0}$ distribution at $0.05\lesssim Y_\mathrm{e}\lesssim0.20$. The survival of the enhancement and depletion after integration therefore indicates a systematic NLD sensitivity across a substantial fraction of the ejecta. 

\subsection{Neutron capture impact on r-process from parametric models}
The previous analysis demonstrates how NLD-dependent $(n, \gamma)$ rates affect the r-process abundance for individual representative trajectories and the total integrated ejecta from hydrodynamical simulations. However, these trajectories sample only a limited range of astrophysical conditions. To systematically investigate the NLD dependence on r-process abundances, we extend the analysis to a grid of parametric trajectories, covering a broad range of initial electron fractions ($0.005\leq Y_\mathrm{e,0}\leq0.5$) and entropies ($1\leq s_0\,[k\mathrm{_B/nuc}]\leq290$) for a typical expansion timescale of $\tau=4\,\mathrm{ms}$, following \citet{Kuske:2025ffp}.
In order to examine the impact of NLD models across the $Y_\mathrm{e,0}$ - $s_0$ parameter space, we use the pairwise abundance difference \citep{Kuske:2025ffp}:
\begin{align}
    \label{eq:massFractionDistance}
    d_{i,j}=\frac{1}{A_\mathrm{max}}\sum_{A=1}^{A_\mathrm{max}}|\log \big(X_i(A)\big)-\log\big(X_j(A)\big)|\,,
\end{align}
i.e., the average logarithmic difference of mass fractions $X_i(A)$ and $X_j(A)$ between two abundance sets $i$ and $j$. We use $A_\mathrm{max}=250$ as the maximum mass number and only consider mass fractions above $10^{-7}$.
This metric is calculated between the LD3, LD4 and reference model (LD5) for each parametric trajectory.

The sensitivity exhibits a strong dependence on astrophysical conditions. As shown in Fig.~\ref{fig:parametric_ld}, the largest abundance differences occur predominantly at high entropy, extending over a broad range of $Y_\mathrm{e,0}$, where the neutron-to-seed ratio at 3\,GK (dashed lines) is extremely large ($\gtrsim1000$). However, these conditions are rarely found in hydrodynamical simulations \citep[cf. Fig.~9 of][]{Kuske:2025ffp}. Two additional regions of enhanced sensitivity are found at intermediate electron fractions ($0.15\lesssim Y_\mathrm{e,0}\lesssim0.30$) at low ($1\lesssim s_0\,[k\mathrm{_B/nuc}]\lesssim 3$) and intermediate ($18\lesssim s_0\,[k\mathrm{_B/nuc}]\lesssim31$) entropies. In contrast, the high $Y_\mathrm{e,0}$, low-to-intermediate entropy region remains largely insensitive to the NLD prescription. Under these conditions, the neutron-to-seed ratio is too small ($\lesssim5$) to drive the reaction flow far from stability, and the nucleosynthesis remains concentrated mainly around the iron-group region. The compared NLDs exhibit broadly the same sensitivity structures, although the magnitude of $d_{i,j}$ differs between the model combinations. This indicates that the location of the sensitive regions is primarily determined by the astrophysical conditions, while the choice of NLD prescriptions controls the magnitude of the resulting abundance variation.

\begin{table}[t]
\caption{Selected trajectories corresponding to the labeled points in Fig.~\ref{fig:parametric_ld}, where $s_0$ and $Y_{e,0}$ denote the initial entropy and electron fraction at $T=7\,\mathrm{GK}$, respectively. For each NLD the time interval $\Delta t=t(\mathrm {RFO})-t(\mathrm{EFO})$ between EFO and RFO is given ($\Delta t>0$ for cold, $\Delta t<0$ for hot r-process).}
\label{tab:points_para}
\begin{tabularx}{\columnwidth}{>{\centering\arraybackslash}X>{\centering\arraybackslash}X>{\centering\arraybackslash}X>{\centering\arraybackslash}X>{\centering\arraybackslash}X>{\centering\arraybackslash}X}
\hline
\hline
Point & $s_0$ & $Y_\mathrm{e,0}$ & $\Delta$t$_\text{LD3}$ & $\Delta$t$_\text{LD4}$ & $\Delta$t$_\text{LD5}$ \\
& [$k\mathrm{_B/nuc}$] & & [ms] &[ms] & [ms] \\
\hline
1 & 22 & 0.245 & 11.5 & 21.3& 7.6\\
2 & 1.95 & 0.240 & -202.3 & -157.2& -170.3\\
3 & 7.75 & 0.210 & 2.5 & 5.8 & 0\\
4 & 185 & 0.435 & 342.0 & 432.4& 309.6\\
5 & 43 & 0.220 & 123.1 & 152.6& 126.6\\
6 & 9.25 & 0.135 & 71.1 & 133.6& 96.9\\
\hline
\hline
\end{tabularx}
\end{table}

We select six conditions (marked in Fig.~\ref{fig:parametric_ld}) for further investigation to highlight the NLD impact across different regions of the parameter space. The selected conditions are commonly found in state-of-the-art hydrodynamical simulations (cf. Fig.~9 of \citep[][]{Kuske:2025ffp}) - except for point 2, which is added due to its unusually strong NLD dependency. Table~\ref{tab:points_para} summarizes the conditions, as well as the time differences between EFO and RFO for the three NLDs. The corresponding final abundances are shown in Fig.~\ref{fig:parametric_interested_points}.

\citet{Kuske:2025ffp} identified eight main nucleosynthesis groups across the parameter space, based on which of the three r-process peaks are formed. Along their boundaries (gray solid lines in Fig.~\ref{fig:parametric_ld}), the impact of NLDs can be particularly strong, because changes in the neutron-capture rates can affect the r-process flow considerably, effectively moving the final abundances from one nucleosynthesis group to another. This boundary effect explains the two curved bands of enhanced sensitivity near points 1 and 2 in Fig.~\ref{fig:parametric_ld}. In those sensitive conditions, the NLDs determine whether the nucleosynthesis flow stops in the second peak or can overcome it. The sensitivity at point 1 is further influenced by a competition between neutron captures and charged-particle reactions \citep[][]{Kuske:2025ffp}, which modifies the r-process path and leads to substantial abundance differences already at RFO.

Points 3 and 4 also lie near the boundaries between nucleosynthesis groups, but exhibit a weaker dependence on the NLD prescription (Fig.~\ref{fig:parametric_interested_points}), especially for the main r-process peaks. However, neutron captures following EFO induce a small shift in the abundances at $A\sim105–120$. In point 4, the Pb abundances ($A\sim205–210$) are strongly impacted by the NLD choice, with an enhancement for LD3 and a depletion for LD4 relative to LD5. This feature originates from differences in the reaction flow through the third r-process peak. LD3 allows more material to reach heavier nuclei before RFO, whereas LD4 reduces the amount of matter around these nuclei relative to the reference calculation. The subsequent decay toward stability preserves this abundance difference, giving rise to the corresponding enhancement and depletion in the final Pb abundances.

Points 5 and 6 are typical cold r-process conditions. For both trajectories, the time between EFO and RFO is longer for LD3 and LD4 relative to LD5 (cf. Table~\ref{tab:points_para}), extending the time period in which neutron captures compete with beta decays. This leads to accumulating differences between the nucleosynthesis of different NLDs, affecting the final abundance patterns, in particular around $A\sim174$ (cf. Section~\ref{sec:impactOfNLDs}).

\section{Conclusions}
\label{sec:conclusions}
In this study, we investigate the impact of statistical model prescriptions on r-process nucleosynthesis using two representative trajectories, mass-integrated ejecta from an NSM simulation, and a broad set of parametric trajectories. We consider all nuclear level density (NLD) and gamma-strength function (GSF) models available in TALYS-2.0 to assess how uncertainties in these ingredients propagate through neutron-capture rates to the final abundances.

We confirm previous findings \citep[see, e.g.,][]{Nikas:2020zcc} in that the final abundances are generally more sensitive to the NLD prescription than to the GSF prescription. For the two considered representative trajectories, most abundance differences are small, resulting in localized enhancements, depletions, or shifts of in the final abundances, while the main r-process peaks remain largely robust. Nevertheless, several pronounced NLD-dependent features emerge. The formation of a subpeak at $A\simeq172-178$ with LD4 and LD6 is associated with neutron-capture flows through Eu and neighboring isotopic chains as the temperature drops to approximately 0.45\,GK, followed by beta decays towards stability. The depletion with LD3 and enhancement with LD4 at $A\simeq180-194$, relative to LD5, are linked to a reversal in the ordering of their neutron-capture rates near the third r-process peak.

We further examine whether these features survive mass integration over a large set of neutron star merger ejecta, since averaging over trajectories with different thermodynamic histories can suppress individual abundance differences. Although some variations are reduced, the LD4 subpeak at $A\simeq172-178$ and the LD3 depletion and LD4 enhancement at $A\simeq180-194$ are preserved in the integrated abundances. The main r-process peaks again remain largely insensitive to the NLD prescription. These results demonstrate that mass integration can reduce trajectory-specific variations, without completely removing systematic effects of the nuclear inputs.

To establish the astrophysical dependence of these sensitivities, we extend the analysis to parametric trajectories that spans a broad range of initial electron fractions $Y_\mathrm{e,0}$ and entropies $s_0$ and covers many conditions found in state-of-the-art hydrodynamical simulations. The resulting abundance differences depend strongly on the thermodynamic conditions. At relatively high electron fractions ($Y_\mathrm{e,0}\gtrsim0.35$) the limited neutron supply restricts heavy element production, and the final abundances generally exhibit less sensitivity to the neutron-capture rates. At lower electron fractions, the sensitivity depends on the nucleosynthesis regime and the relative timing of equilibrium freeze-out (EFO) and r-process freeze-out (RFO). In the hot r-process conditions encountered at lower entropies, EFO occurs after RFO. By the time photodissociations become subdominant to neutron captures, free neutrons are largely exhausted, limiting the subsequent redistribution of material by NLD-dependent capture rates. The final abundances therefore generally show smaller differences. Nevertheless, pronounced variations occur near the boundaries between nucleosynthesis groups, where minor changes in reaction rates can alter the resulting abundance pattern substantially. Such boundary effects are also present in the cold r-process regime, which are found at low $Y_\mathrm{e,0}$ and high entropy. In these conditions, EFO occurs before RFO, i.e., some free neutrons remain after photodissociations have become ineffective. Post-freezeout neutron captures can therefore reshape the abundance distribution, resulting in a stronger sensitivity to the NLD prescription.

Overall, our results show that an accurate understanding of NLDs is crucial for reliable r-process nucleosynthesis predictions. They can considerably alter neutron-capture rates, leading to differences in the post-freezeout redistribution of abundances, in particular under cold r-process conditions. For some sensitive conditions, the NLDs determine whether a shell closure can be overcome, i.e., whether the second or third peak is reached. Finally, NLDs also impact the nuclear partition functions, leading to minor differences already in the initial abundances and during $(n,\gamma)-(\gamma,n)$ equilibrium. A combination of experimental and theoretical efforts is required to decrease the remaining NLD model uncertainties.

\begin{acknowledgments}
We thank Johann Isaak, Gail McLaughlin, and Stefan Typel for useful discussions. This work was supported by the Helmholtz Forschungsakademie Hessen für FAIR (HFHF) and Deutsche Forschungsgemeinschaft (DFG, German Research Foundation) -- Project-ID 279384907 -- SFB 1245. 
Calculations for this research were conducted on the Lichtenberg high performance computer of the TU Darmstadt (Project-ID l0003035 and l0003406). \\
\end{acknowledgments}

\section*{Data Availability}
The neutron-capture rates and nucleosynthesis calculation data supporting the findings of this study are available upon reasonable request.

\bibliography{apssamp}

\begin{thebibliography}{93}%
\makeatletter
\providecommand \@ifxundefined [1]{%
 \@ifx{#1\undefined}
}%
\providecommand \@ifnum [1]{%
 \ifnum #1\expandafter \@firstoftwo
 \else \expandafter \@secondoftwo
 \fi
}%
\providecommand \@ifx [1]{%
 \ifx #1\expandafter \@firstoftwo
 \else \expandafter \@secondoftwo
 \fi
}%
\providecommand \natexlab [1]{#1}%
\providecommand \enquote  [1]{``#1''}%
\providecommand \bibnamefont  [1]{#1}%
\providecommand \bibfnamefont [1]{#1}%
\providecommand \citenamefont [1]{#1}%
\providecommand \href@noop [0]{\@secondoftwo}%
\providecommand \href [0]{\begingroup \@sanitize@url \@href}%
\providecommand \@href[1]{\@@startlink{#1}\@@href}%
\providecommand \@@href[1]{\endgroup#1\@@endlink}%
\providecommand \@sanitize@url [0]{\catcode `\\12\catcode `\$12\catcode `\&12\catcode `\#12\catcode `\^12\catcode `\_12\catcode `\%12\relax}%
\providecommand \@@startlink[1]{}%
\providecommand \@@endlink[0]{}%
\providecommand \url  [0]{\begingroup\@sanitize@url \@url }%
\providecommand \@url [1]{\endgroup\@href {#1}{\urlprefix }}%
\providecommand \urlprefix  [0]{URL }%
\providecommand \Eprint [0]{\href }%
\providecommand \doibase [0]{https://doi.org/}%
\providecommand \selectlanguage [0]{\@gobble}%
\providecommand \bibinfo  [0]{\@secondoftwo}%
\providecommand \bibfield  [0]{\@secondoftwo}%
\providecommand \translation [1]{[#1]}%
\providecommand \BibitemOpen [0]{}%
\providecommand \bibitemStop [0]{}%
\providecommand \bibitemNoStop [0]{.\EOS\space}%
\providecommand \EOS [0]{\spacefactor3000\relax}%
\providecommand \BibitemShut  [1]{\csname bibitem#1\endcsname}%
\let\auto@bib@innerbib\@empty
\bibitem [{\citenamefont {Arnould}\ \emph {et~al.}(2007)\citenamefont {Arnould}, \citenamefont {Goriely},\ and\ \citenamefont {Takahashi}}]{Arnould2007_RprocessStellarNucleosynthesis}%
  \BibitemOpen
  \bibfield  {author} {\bibinfo {author} {\bibfnamefont {M.}~\bibnamefont {Arnould}}, \bibinfo {author} {\bibfnamefont {S.}~\bibnamefont {Goriely}},\ and\ \bibinfo {author} {\bibfnamefont {K.}~\bibnamefont {Takahashi}},\ }\bibfield  {title} {\bibinfo {title} {The r-process of stellar nucleosynthesis: {{Astrophysics}} and nuclear physics achievements and mysteries},\ }\href {https://doi.org/10.1016/j.physrep.2007.06.002} {\bibfield  {journal} {\bibinfo  {journal} {Phys. Rept.}\ }\textbf {\bibinfo {volume} {450}},\ \bibinfo {pages} {97} (\bibinfo {year} {2007})}\BibitemShut {NoStop}%
\bibitem [{\citenamefont {Horowitz}\ \emph {et~al.}(2019)\citenamefont {Horowitz} \emph {et~al.}}]{Horowitz:2018ndv}%
  \BibitemOpen
  \bibfield  {author} {\bibinfo {author} {\bibfnamefont {C.~J.}\ \bibnamefont {Horowitz}} \emph {et~al.},\ }\bibfield  {title} {\bibinfo {title} {{$r$-Process Nucleosynthesis: Connecting Rare-Isotope Beam Facilities with the Cosmos}},\ }\href {https://doi.org/10.1088/1361-6471/ab0849} {\bibfield  {journal} {\bibinfo  {journal} {J. Phys. G}\ }\textbf {\bibinfo {volume} {46}},\ \bibinfo {pages} {083001} (\bibinfo {year} {2019})}\BibitemShut {NoStop}%
\bibitem [{\citenamefont {Cowan}\ \emph {et~al.}(2021)\citenamefont {Cowan}, \citenamefont {Sneden}, \citenamefont {Lawler}, \citenamefont {Aprahamian}, \citenamefont {Wiescher}, \citenamefont {Langanke}, \citenamefont {Mart{\'\i}nez-Pinedo},\ and\ \citenamefont {Thielemann}}]{Cowan:2019pkx}%
  \BibitemOpen
  \bibfield  {author} {\bibinfo {author} {\bibfnamefont {J.~J.}\ \bibnamefont {Cowan}}, \bibinfo {author} {\bibfnamefont {C.}~\bibnamefont {Sneden}}, \bibinfo {author} {\bibfnamefont {J.~E.}\ \bibnamefont {Lawler}}, \bibinfo {author} {\bibfnamefont {A.}~\bibnamefont {Aprahamian}}, \bibinfo {author} {\bibfnamefont {M.}~\bibnamefont {Wiescher}}, \bibinfo {author} {\bibfnamefont {K.}~\bibnamefont {Langanke}}, \bibinfo {author} {\bibfnamefont {G.}~\bibnamefont {Mart{\'\i}nez-Pinedo}},\ and\ \bibinfo {author} {\bibfnamefont {F.-K.}\ \bibnamefont {Thielemann}},\ }\bibfield  {title} {\bibinfo {title} {{Origin of the heaviest elements: The rapid neutron-capture process}},\ }\href {https://doi.org/10.1103/RevModPhys.93.015002} {\bibfield  {journal} {\bibinfo  {journal} {Rev. Mod. Phys.}\ }\textbf {\bibinfo {volume} {93}},\ \bibinfo {pages} {15002} (\bibinfo {year} {2021})}\BibitemShut {NoStop}%
\bibitem [{\citenamefont {Arcones}\ and\ \citenamefont {Thielemann}(2023)}]{Arcones:2022jer}%
  \BibitemOpen
  \bibfield  {author} {\bibinfo {author} {\bibfnamefont {A.}~\bibnamefont {Arcones}}\ and\ \bibinfo {author} {\bibfnamefont {F.-K.}\ \bibnamefont {Thielemann}},\ }\bibfield  {title} {\bibinfo {title} {{Origin of the elements}},\ }\href {https://doi.org/10.1007/s00159-022-00146-x} {\bibfield  {journal} {\bibinfo  {journal} {Astron. Astrophys. Rev.}\ }\textbf {\bibinfo {volume} {31}},\ \bibinfo {pages} {1} (\bibinfo {year} {2023})}\BibitemShut {NoStop}%
\bibitem [{\citenamefont {{Thielemann}}\ and\ \citenamefont {{Cowan}}(2026)}]{2026EPJA...62..105T}%
  \BibitemOpen
  \bibfield  {author} {\bibinfo {author} {\bibfnamefont {F.-K.}\ \bibnamefont {{Thielemann}}}\ and\ \bibinfo {author} {\bibfnamefont {J.~J.}\ \bibnamefont {{Cowan}}},\ }\bibfield  {title} {\bibinfo {title} {{The r-process: history, required conditions, astrophysical sites, and observations}},\ }\href {https://doi.org/10.1140/epja/s10050-026-01865-w} {\bibfield  {journal} {\bibinfo  {journal} {European Phys. J. A}\ }\textbf {\bibinfo {volume} {62}},\ \bibinfo {eid} {105} (\bibinfo {year} {2026})}\BibitemShut {NoStop}%
\bibitem [{\citenamefont {Abbott}\ \emph {et~al.}(2017)\citenamefont {Abbott} \emph {et~al.}}]{Abbott2017_MultimessengerObservationsBinary}%
  \BibitemOpen
  \bibfield  {author} {\bibinfo {author} {\bibfnamefont {B.~P.}\ \bibnamefont {Abbott}} \emph {et~al.} (\bibinfo {collaboration} {LIGO Scientific et al.}),\ }\bibfield  {title} {\bibinfo {title} {Multi-messenger observations of a binary neutron star merger},\ }\href {https://doi.org/10.3847/2041-8213/aa91c9} {\bibfield  {journal} {\bibinfo  {journal} {Astrophys. J. Lett.}\ }\textbf {\bibinfo {volume} {848}},\ \bibinfo {pages} {L12} (\bibinfo {year} {2017})}\BibitemShut {NoStop}%
\bibitem [{\citenamefont {Drout}\ \emph {et~al.}(2017)\citenamefont {Drout} \emph {et~al.}}]{Drout2017_LightCurvesNeutron}%
  \BibitemOpen
  \bibfield  {author} {\bibinfo {author} {\bibfnamefont {M.~R.}\ \bibnamefont {Drout}} \emph {et~al.},\ }\bibfield  {title} {\bibinfo {title} {Light curves of the neutron star merger {{GW170817}}/{{SSS17a}}: {{Implications}} for r-process nucleosynthesis},\ }\href {https://doi.org/10.1126/science.aaq0049} {\bibfield  {journal} {\bibinfo  {journal} {Science}\ }\textbf {\bibinfo {volume} {358}},\ \bibinfo {pages} {1570} (\bibinfo {year} {2017})}\BibitemShut {NoStop}%
\bibitem [{\citenamefont {Kilpatrick}\ \emph {et~al.}(2017)\citenamefont {Kilpatrick} \emph {et~al.}}]{Kilpatrick2017_ElectromagneticEvidenceThat}%
  \BibitemOpen
  \bibfield  {author} {\bibinfo {author} {\bibfnamefont {C.~D.}\ \bibnamefont {Kilpatrick}} \emph {et~al.},\ }\bibfield  {title} {\bibinfo {title} {Electromagnetic evidence that {{SSS17a}} is the result of a binary neutron star merger},\ }\href {https://doi.org/10.1126/science.aaq0073} {\bibfield  {journal} {\bibinfo  {journal} {Science}\ }\textbf {\bibinfo {volume} {358}},\ \bibinfo {pages} {1583} (\bibinfo {year} {2017})}\BibitemShut {NoStop}%
\bibitem [{\citenamefont {Argast}\ \emph {et~al.}(2004)\citenamefont {Argast}, \citenamefont {Samland}, \citenamefont {Thielemann},\ and\ \citenamefont {Qian}}]{Argast2004_NeutronStarMergers}%
  \BibitemOpen
  \bibfield  {author} {\bibinfo {author} {\bibfnamefont {D.}~\bibnamefont {Argast}}, \bibinfo {author} {\bibfnamefont {M.}~\bibnamefont {Samland}}, \bibinfo {author} {\bibfnamefont {F.-K.}\ \bibnamefont {Thielemann}},\ and\ \bibinfo {author} {\bibfnamefont {Y.-Z.}\ \bibnamefont {Qian}},\ }\bibfield  {title} {\bibinfo {title} {Neutron star mergers versus core-collapse supernovae as dominant r-process sites in the early {{Galaxy}}},\ }\href {https://doi.org/10.1051/0004-6361:20034265} {\bibfield  {journal} {\bibinfo  {journal} {Astron. \& Astrophys.}\ }\textbf {\bibinfo {volume} {416}},\ \bibinfo {pages} {997} (\bibinfo {year} {2004})}\BibitemShut {NoStop}%
\bibitem [{\citenamefont {{Matteucci}}\ \emph {et~al.}(2014)\citenamefont {{Matteucci}}, \citenamefont {{Romano}}, \citenamefont {{Arcones}}, \citenamefont {{Korobkin}},\ and\ \citenamefont {{Rosswog}}}]{2014MNRAS.438.2177M}%
  \BibitemOpen
  \bibfield  {author} {\bibinfo {author} {\bibfnamefont {F.}~\bibnamefont {{Matteucci}}}, \bibinfo {author} {\bibfnamefont {D.}~\bibnamefont {{Romano}}}, \bibinfo {author} {\bibfnamefont {A.}~\bibnamefont {{Arcones}}}, \bibinfo {author} {\bibfnamefont {O.}~\bibnamefont {{Korobkin}}},\ and\ \bibinfo {author} {\bibfnamefont {S.}~\bibnamefont {{Rosswog}}},\ }\bibfield  {title} {\bibinfo {title} {{Europium production: neutron star mergers versus core-collapse supernovae}},\ }\href {https://doi.org/10.1093/mnras/stt2350} {\bibfield  {journal} {\bibinfo  {journal} {Mon. Not. Roy. Astron. Soc.}\ }\textbf {\bibinfo {volume} {438}},\ \bibinfo {pages} {2177} (\bibinfo {year} {2014})}\BibitemShut {NoStop}%
\bibitem [{\citenamefont {C{\^o}t{\'e}}\ \emph {et~al.}(2019)\citenamefont {C{\^o}t{\'e}}, \citenamefont {Eichler}, \citenamefont {Arcones}, \citenamefont {Hansen}, \citenamefont {Simonetti}, \citenamefont {Frebel}, \citenamefont {Fryer}, \citenamefont {Pignatari}, \citenamefont {Reichert}, \citenamefont {Belczynski},\ and\ \citenamefont {Matteucci}}]{Cote2019_NeutronStarMergers}%
  \BibitemOpen
  \bibfield  {author} {\bibinfo {author} {\bibfnamefont {B.}~\bibnamefont {C{\^o}t{\'e}}}, \bibinfo {author} {\bibfnamefont {M.}~\bibnamefont {Eichler}}, \bibinfo {author} {\bibfnamefont {A.}~\bibnamefont {Arcones}}, \bibinfo {author} {\bibfnamefont {C.~J.}\ \bibnamefont {Hansen}}, \bibinfo {author} {\bibfnamefont {P.}~\bibnamefont {Simonetti}}, \bibinfo {author} {\bibfnamefont {A.}~\bibnamefont {Frebel}}, \bibinfo {author} {\bibfnamefont {C.~L.}\ \bibnamefont {Fryer}}, \bibinfo {author} {\bibfnamefont {M.}~\bibnamefont {Pignatari}}, \bibinfo {author} {\bibfnamefont {M.}~\bibnamefont {Reichert}}, \bibinfo {author} {\bibfnamefont {K.}~\bibnamefont {Belczynski}},\ and\ \bibinfo {author} {\bibfnamefont {F.}~\bibnamefont {Matteucci}},\ }\bibfield  {title} {\bibinfo {title} {Neutron {{Star Mergers Might Not Be}} the {{Only Source}} of r-process {{Elements}} in the {{Milky Way}}},\ }\href {https://doi.org/10.3847/1538-4357/ab10db} {\bibfield  {journal} {\bibinfo  {journal} {Astrophys. J.}\ }\textbf {\bibinfo
  {volume} {875}},\ \bibinfo {pages} {106} (\bibinfo {year} {2019})}\BibitemShut {NoStop}%
\bibitem [{\citenamefont {Molero}\ \emph {et~al.}(2023)\citenamefont {Molero}, \citenamefont {Magrini}, \citenamefont {Matteucci}, \citenamefont {Romano}, \citenamefont {Palla}, \citenamefont {Cescutti}, \citenamefont {V{\'a}zquez},\ and\ \citenamefont {Spitoni}}]{Molero2023_OriginNeutronCapture}%
  \BibitemOpen
  \bibfield  {author} {\bibinfo {author} {\bibfnamefont {M.}~\bibnamefont {Molero}}, \bibinfo {author} {\bibfnamefont {L.}~\bibnamefont {Magrini}}, \bibinfo {author} {\bibfnamefont {F.}~\bibnamefont {Matteucci}}, \bibinfo {author} {\bibfnamefont {D.}~\bibnamefont {Romano}}, \bibinfo {author} {\bibfnamefont {M.}~\bibnamefont {Palla}}, \bibinfo {author} {\bibfnamefont {G.}~\bibnamefont {Cescutti}}, \bibinfo {author} {\bibfnamefont {C.~V.}\ \bibnamefont {V{\'a}zquez}},\ and\ \bibinfo {author} {\bibfnamefont {E.}~\bibnamefont {Spitoni}},\ }\bibfield  {title} {\bibinfo {title} {Origin of neutron capture elements with the {{Gaia-ESO}} survey: The evolution of s- and r-process elements across the {{Milky Way}}},\ }\href {https://doi.org/10.1093/mnras/stad1577} {\bibfield  {journal} {\bibinfo  {journal} {Mon. Not. Roy. Astron. Soc.}\ }\textbf {\bibinfo {volume} {523}},\ \bibinfo {pages} {2974} (\bibinfo {year} {2023})}\BibitemShut {NoStop}%
\bibitem [{\citenamefont {Winteler}\ \emph {et~al.}(2012)\citenamefont {Winteler}, \citenamefont {K{\"a}ppeli}, \citenamefont {Perego}, \citenamefont {Arcones}, \citenamefont {Vasset}, \citenamefont {Nishimura}, \citenamefont {Liebend{\"o}rfer},\ and\ \citenamefont {Thielemann}}]{Winteler2012_MagnetorotationallyDrivenSupernovae}%
  \BibitemOpen
  \bibfield  {author} {\bibinfo {author} {\bibfnamefont {C.}~\bibnamefont {Winteler}}, \bibinfo {author} {\bibfnamefont {R.}~\bibnamefont {K{\"a}ppeli}}, \bibinfo {author} {\bibfnamefont {A.}~\bibnamefont {Perego}}, \bibinfo {author} {\bibfnamefont {A.}~\bibnamefont {Arcones}}, \bibinfo {author} {\bibfnamefont {N.}~\bibnamefont {Vasset}}, \bibinfo {author} {\bibfnamefont {N.}~\bibnamefont {Nishimura}}, \bibinfo {author} {\bibfnamefont {M.}~\bibnamefont {Liebend{\"o}rfer}},\ and\ \bibinfo {author} {\bibfnamefont {F.-K.}\ \bibnamefont {Thielemann}},\ }\bibfield  {title} {\bibinfo {title} {Magnetorotationally {{Driven Supernovae}} as the {{Origin}} of {{Early Galaxy R-Process Elements}}?},\ }\href {https://doi.org/10.1088/2041-8205/750/1/L22} {\bibfield  {journal} {\bibinfo  {journal} {Astrophys. J. Lett.}\ }\textbf {\bibinfo {volume} {750}},\ \bibinfo {pages} {L22} (\bibinfo {year} {2012})}\BibitemShut {NoStop}%
\bibitem [{\citenamefont {Reichert}\ \emph {et~al.}(2021)\citenamefont {Reichert}, \citenamefont {Obergaulinger}, \citenamefont {Eichler}, \citenamefont {Aloy},\ and\ \citenamefont {Arcones}}]{Reichert2021_NucleosynthesisMagnetorotationalSupernovae}%
  \BibitemOpen
  \bibfield  {author} {\bibinfo {author} {\bibfnamefont {M.}~\bibnamefont {Reichert}}, \bibinfo {author} {\bibfnamefont {M.}~\bibnamefont {Obergaulinger}}, \bibinfo {author} {\bibfnamefont {M.}~\bibnamefont {Eichler}}, \bibinfo {author} {\bibfnamefont {M.~{\'A}.}\ \bibnamefont {Aloy}},\ and\ \bibinfo {author} {\bibfnamefont {A.}~\bibnamefont {Arcones}},\ }\bibfield  {title} {\bibinfo {title} {Nucleosynthesis in magneto-rotational supernovae},\ }\href {https://doi.org/10.1093/mnras/stab029} {\bibfield  {journal} {\bibinfo  {journal} {Mon. Not. Roy. Astron. Soc.}\ }\textbf {\bibinfo {volume} {501}},\ \bibinfo {pages} {5733} (\bibinfo {year} {2021})}\BibitemShut {NoStop}%
\bibitem [{\citenamefont {Siegel}\ \emph {et~al.}(2019)\citenamefont {Siegel}, \citenamefont {Barnes},\ and\ \citenamefont {Metzger}}]{Siegel2019_CollapsarsMajorSource}%
  \BibitemOpen
  \bibfield  {author} {\bibinfo {author} {\bibfnamefont {D.~M.}\ \bibnamefont {Siegel}}, \bibinfo {author} {\bibfnamefont {J.}~\bibnamefont {Barnes}},\ and\ \bibinfo {author} {\bibfnamefont {B.~D.}\ \bibnamefont {Metzger}},\ }\bibfield  {title} {\bibinfo {title} {Collapsars as a major source of r-process elements},\ }\href {https://doi.org/10.1038/s41586-019-1136-0} {\bibfield  {journal} {\bibinfo  {journal} {Nature}\ }\textbf {\bibinfo {volume} {569}},\ \bibinfo {pages} {241} (\bibinfo {year} {2019})}\BibitemShut {NoStop}%
\bibitem [{\citenamefont {Patel}\ \emph {et~al.}(2025)\citenamefont {Patel}, \citenamefont {Metzger}, \citenamefont {Cehula}, \citenamefont {Burns}, \citenamefont {Goldberg},\ and\ \citenamefont {Thompson}}]{Patel2025_DirectEvidenceRprocess}%
  \BibitemOpen
  \bibfield  {author} {\bibinfo {author} {\bibfnamefont {A.}~\bibnamefont {Patel}}, \bibinfo {author} {\bibfnamefont {B.~D.}\ \bibnamefont {Metzger}}, \bibinfo {author} {\bibfnamefont {J.}~\bibnamefont {Cehula}}, \bibinfo {author} {\bibfnamefont {E.}~\bibnamefont {Burns}}, \bibinfo {author} {\bibfnamefont {J.~A.}\ \bibnamefont {Goldberg}},\ and\ \bibinfo {author} {\bibfnamefont {T.~A.}\ \bibnamefont {Thompson}},\ }\bibfield  {title} {\bibinfo {title} {Direct {{Evidence}} for r-process {{Nucleosynthesis}} in {{Delayed MeV Emission}} from the {{SGR}} 1806--20 {{Magnetar Giant Flare}}},\ }\href {https://doi.org/10.3847/2041-8213/adc9b0} {\bibfield  {journal} {\bibinfo  {journal} {Astrophys. J. Lett.}\ }\textbf {\bibinfo {volume} {984}},\ \bibinfo {pages} {L29} (\bibinfo {year} {2025})}\BibitemShut {NoStop}%
\bibitem [{\citenamefont {Kratz}\ \emph {et~al.}(1993)\citenamefont {Kratz}, \citenamefont {Bitouzet}, \citenamefont {Thielemann}, \citenamefont {Moeller},\ and\ \citenamefont {Pfeiffer}}]{Kratz1993_IsotopicRProcessAbundances}%
  \BibitemOpen
  \bibfield  {author} {\bibinfo {author} {\bibfnamefont {K.-L.}\ \bibnamefont {Kratz}}, \bibinfo {author} {\bibfnamefont {J.-P.}\ \bibnamefont {Bitouzet}}, \bibinfo {author} {\bibfnamefont {F.-K.}\ \bibnamefont {Thielemann}}, \bibinfo {author} {\bibfnamefont {P.}~\bibnamefont {Moeller}},\ and\ \bibinfo {author} {\bibfnamefont {B.}~\bibnamefont {Pfeiffer}},\ }\bibfield  {title} {\bibinfo {title} {Isotopic r-{{Process}} abundances and nuclear structure far from stability: {{Implications}} for the r-{{Process}} mechanism},\ }\href {https://doi.org/10.1086/172196} {\bibfield  {journal} {\bibinfo  {journal} {Astrophys. J.}\ }\textbf {\bibinfo {volume} {403}},\ \bibinfo {pages} {216} (\bibinfo {year} {1993})}\BibitemShut {NoStop}%
\bibitem [{\citenamefont {Goriely}(2001)}]{Goriely2001_NuclearMassesPProcesses}%
  \BibitemOpen
  \bibfield  {author} {\bibinfo {author} {\bibfnamefont {S.}~\bibnamefont {Goriely}},\ }\bibfield  {title} {\bibinfo {title} {Nuclear {{Masses}} and the r- and p-{{Processes}} of {{Nucleosynthesis}}},\ }in\ \href {https://doi.org/10.1007/978-94-015-1270-1_5} {\emph {\bibinfo {booktitle} {Atomic Physics at Accelerators: Mass Spectrometry}}}\ (\bibinfo {year} {2001})\BibitemShut {NoStop}%
\bibitem [{\citenamefont {Wanajo}\ \emph {et~al.}(2004)\citenamefont {Wanajo}, \citenamefont {Goriely}, \citenamefont {Samyn},\ and\ \citenamefont {Itoh}}]{Wanajo:2004ai}%
  \BibitemOpen
  \bibfield  {author} {\bibinfo {author} {\bibfnamefont {S.}~\bibnamefont {Wanajo}}, \bibinfo {author} {\bibfnamefont {S.}~\bibnamefont {Goriely}}, \bibinfo {author} {\bibfnamefont {M.}~\bibnamefont {Samyn}},\ and\ \bibinfo {author} {\bibfnamefont {N.}~\bibnamefont {Itoh}},\ }\bibfield  {title} {\bibinfo {title} {{The r-Process in supernovae: Impact of new microscopic mass formulas}},\ }\href {https://doi.org/10.1086/383140} {\bibfield  {journal} {\bibinfo  {journal} {Astrophys. J.}\ }\textbf {\bibinfo {volume} {606}},\ \bibinfo {pages} {1057} (\bibinfo {year} {2004})}\BibitemShut {NoStop}%
\bibitem [{\citenamefont {Arcones}\ and\ \citenamefont {Martinez-Pinedo}(2011)}]{Arcones:2010dz}%
  \BibitemOpen
  \bibfield  {author} {\bibinfo {author} {\bibfnamefont {A.}~\bibnamefont {Arcones}}\ and\ \bibinfo {author} {\bibfnamefont {G.}~\bibnamefont {Martinez-Pinedo}},\ }\bibfield  {title} {\bibinfo {title} {{Dynamical r-process studies within the neutrino-driven wind scenario and its sensitivity to the nuclear physics input}},\ }\href {https://doi.org/10.1103/PhysRevC.83.045809} {\bibfield  {journal} {\bibinfo  {journal} {Phys. Rev. C}\ }\textbf {\bibinfo {volume} {83}},\ \bibinfo {pages} {045809} (\bibinfo {year} {2011})}\BibitemShut {NoStop}%
\bibitem [{\citenamefont {Mumpower}\ \emph {et~al.}(2015)\citenamefont {Mumpower}, \citenamefont {Surman}, \citenamefont {Fang}, \citenamefont {Beard}, \citenamefont {M{\"o}ller}, \citenamefont {Kawano},\ and\ \citenamefont {Aprahamian}}]{Mumpower:2015hva}%
  \BibitemOpen
  \bibfield  {author} {\bibinfo {author} {\bibfnamefont {M.~R.}\ \bibnamefont {Mumpower}}, \bibinfo {author} {\bibfnamefont {R.}~\bibnamefont {Surman}}, \bibinfo {author} {\bibfnamefont {D.~L.}\ \bibnamefont {Fang}}, \bibinfo {author} {\bibfnamefont {M.}~\bibnamefont {Beard}}, \bibinfo {author} {\bibfnamefont {P.}~\bibnamefont {M{\"o}ller}}, \bibinfo {author} {\bibfnamefont {T.}~\bibnamefont {Kawano}},\ and\ \bibinfo {author} {\bibfnamefont {A.}~\bibnamefont {Aprahamian}},\ }\bibfield  {title} {\bibinfo {title} {{Impact of individual nuclear masses on r-process abundances}},\ }\href {https://doi.org/10.1103/PhysRevC.92.035807} {\bibfield  {journal} {\bibinfo  {journal} {Phys. Rev. C}\ }\textbf {\bibinfo {volume} {92}},\ \bibinfo {pages} {035807} (\bibinfo {year} {2015})}\BibitemShut {NoStop}%
\bibitem [{\citenamefont {Martin}\ \emph {et~al.}(2016)\citenamefont {Martin}, \citenamefont {Arcones}, \citenamefont {Nazarewicz},\ and\ \citenamefont {Olsen}}]{Martin:2015xql}%
  \BibitemOpen
  \bibfield  {author} {\bibinfo {author} {\bibfnamefont {D.}~\bibnamefont {Martin}}, \bibinfo {author} {\bibfnamefont {A.}~\bibnamefont {Arcones}}, \bibinfo {author} {\bibfnamefont {W.}~\bibnamefont {Nazarewicz}},\ and\ \bibinfo {author} {\bibfnamefont {E.}~\bibnamefont {Olsen}},\ }\bibfield  {title} {\bibinfo {title} {{Impact of nuclear mass uncertainties on the $r$-process}},\ }\href {https://doi.org/10.1103/PhysRevLett.116.121101} {\bibfield  {journal} {\bibinfo  {journal} {Phys. Rev. Lett.}\ }\textbf {\bibinfo {volume} {116}},\ \bibinfo {pages} {121101} (\bibinfo {year} {2016})}\BibitemShut {NoStop}%
\bibitem [{\citenamefont {de~Jes{\'u}s Mendoza-Temis}\ \emph {et~al.}(2015)\citenamefont {de~Jes{\'u}s Mendoza-Temis}, \citenamefont {Wu}, \citenamefont {Mart{\'\i}nez-Pinedo}, \citenamefont {Langanke}, \citenamefont {Bauswein},\ and\ \citenamefont {Janka}}]{deJesusMendoza-Temis:2014owk}%
  \BibitemOpen
  \bibfield  {author} {\bibinfo {author} {\bibfnamefont {J.}~\bibnamefont {de~Jes{\'u}s Mendoza-Temis}}, \bibinfo {author} {\bibfnamefont {M.-R.}\ \bibnamefont {Wu}}, \bibinfo {author} {\bibfnamefont {G.}~\bibnamefont {Mart{\'\i}nez-Pinedo}}, \bibinfo {author} {\bibfnamefont {K.}~\bibnamefont {Langanke}}, \bibinfo {author} {\bibfnamefont {A.}~\bibnamefont {Bauswein}},\ and\ \bibinfo {author} {\bibfnamefont {H.-T.}\ \bibnamefont {Janka}},\ }\bibfield  {title} {\bibinfo {title} {{Nuclear robustness of the r-process in neutron-star mergers}},\ }\href {https://doi.org/10.1103/PhysRevC.92.055805} {\bibfield  {journal} {\bibinfo  {journal} {Phys. Rev. C}\ }\textbf {\bibinfo {volume} {92}},\ \bibinfo {pages} {055805} (\bibinfo {year} {2015})}\BibitemShut {NoStop}%
\bibitem [{\citenamefont {Vassh}\ \emph {et~al.}(2021)\citenamefont {Vassh}, \citenamefont {McLaughlin}, \citenamefont {Mumpower},\ and\ \citenamefont {Surman}}]{Vassh2021_MarkovChainMonte}%
  \BibitemOpen
  \bibfield  {author} {\bibinfo {author} {\bibfnamefont {N.}~\bibnamefont {Vassh}}, \bibinfo {author} {\bibfnamefont {G.~C.}\ \bibnamefont {McLaughlin}}, \bibinfo {author} {\bibfnamefont {M.~R.}\ \bibnamefont {Mumpower}},\ and\ \bibinfo {author} {\bibfnamefont {R.}~\bibnamefont {Surman}},\ }\bibfield  {title} {\bibinfo {title} {Markov {{Chain Monte Carlo Predictions}} of {{Neutron-rich Lanthanide Properties}} as a {{Probe}} of r-process {{Dynamics}}},\ }\href {https://doi.org/10.3847/1538-4357/abd035} {\bibfield  {journal} {\bibinfo  {journal} {Astrophys. J.}\ }\textbf {\bibinfo {volume} {907}},\ \bibinfo {pages} {98} (\bibinfo {year} {2021})}\BibitemShut {NoStop}%
\bibitem [{\citenamefont {Giuliani}\ \emph {et~al.}(2026)\citenamefont {Giuliani}, \citenamefont {Mart{\'\i}nez-Pinedo}, \citenamefont {Bauswein},\ and\ \citenamefont {Vijayan}}]{Giuliani:2024ppp}%
  \BibitemOpen
  \bibfield  {author} {\bibinfo {author} {\bibfnamefont {S.~A.}\ \bibnamefont {Giuliani}}, \bibinfo {author} {\bibfnamefont {G.}~\bibnamefont {Mart{\'\i}nez-Pinedo}}, \bibinfo {author} {\bibfnamefont {A.}~\bibnamefont {Bauswein}},\ and\ \bibinfo {author} {\bibfnamefont {V.}~\bibnamefont {Vijayan}},\ }\bibfield  {title} {\bibinfo {title} {{Impact of nuclear masses on r-process nucleosynthesis: Bulk properties versus shell effects}},\ }\href {https://doi.org/10.1103/k64z-32sr} {\bibfield  {journal} {\bibinfo  {journal} {Phys. Rev. C}\ }\textbf {\bibinfo {volume} {113}},\ \bibinfo {pages} {045806} (\bibinfo {year} {2026})}\BibitemShut {NoStop}%
\bibitem [{\citenamefont {Kuske}\ \emph {et~al.}(2026)\citenamefont {Kuske}, \citenamefont {Miyagi}, \citenamefont {Arcones},\ and\ \citenamefont {Schwenk}}]{Kuske:2025tsm}%
  \BibitemOpen
  \bibfield  {author} {\bibinfo {author} {\bibfnamefont {J.}~\bibnamefont {Kuske}}, \bibinfo {author} {\bibfnamefont {T.}~\bibnamefont {Miyagi}}, \bibinfo {author} {\bibfnamefont {A.}~\bibnamefont {Arcones}},\ and\ \bibinfo {author} {\bibfnamefont {A.}~\bibnamefont {Schwenk}},\ }\bibfield  {title} {\bibinfo {title} {{r-Process Nucleosynthesis with Ab~Initio Nuclear Masses around the $N=82$ Shell Closure}},\ }\href {https://doi.org/10.1103/k4lv-436c} {\bibfield  {journal} {\bibinfo  {journal} {Phys. Rev. Lett.}\ }\textbf {\bibinfo {volume} {137}},\ \bibinfo {pages} {052701} (\bibinfo {year} {2026})}\BibitemShut {NoStop}%
\bibitem [{\citenamefont {Martinet}\ and\ \citenamefont {Goriely}(2025)}]{Martinet2025_ImpactMassUncertainties}%
  \BibitemOpen
  \bibfield  {author} {\bibinfo {author} {\bibfnamefont {S.}~\bibnamefont {Martinet}}\ and\ \bibinfo {author} {\bibfnamefont {S.}~\bibnamefont {Goriely}},\ }\bibfield  {title} {\bibinfo {title} {The impact of mass uncertainties on $r$-process nucleosynthesis in neutron star mergers},\ }\href {https://doi.org/10.1051/0004-6361/202451991} {\bibfield  {journal} {\bibinfo  {journal} {Astron. Astrophys.}\ }\textbf {\bibinfo {volume} {694}},\ \bibinfo {pages} {A180} (\bibinfo {year} {2025})}\BibitemShut {NoStop}%
\bibitem [{\citenamefont {Marketin}\ \emph {et~al.}(2016)\citenamefont {Marketin}, \citenamefont {Huther},\ and\ \citenamefont {{Mart{\'i}nez-Pinedo}}}]{Marketin2016_LargescaleEvaluation$ensuremathbeta$decay}%
  \BibitemOpen
  \bibfield  {author} {\bibinfo {author} {\bibfnamefont {T.}~\bibnamefont {Marketin}}, \bibinfo {author} {\bibfnamefont {L.}~\bibnamefont {Huther}},\ and\ \bibinfo {author} {\bibfnamefont {G.}~\bibnamefont {{Mart{\'i}nez-Pinedo}}},\ }\bibfield  {title} {\bibinfo {title} {Large-scale evaluation of $\beta$-decay rates of $r$-process nuclei with the inclusion of first-forbidden transitions},\ }\href {https://doi.org/10.1103/PhysRevC.93.025805} {\bibfield  {journal} {\bibinfo  {journal} {Phys. Rev. C}\ }\textbf {\bibinfo {volume} {93}},\ \bibinfo {pages} {025805} (\bibinfo {year} {2016})}\BibitemShut {NoStop}%
\bibitem [{\citenamefont {Lund}\ \emph {et~al.}(2023)\citenamefont {Lund}, \citenamefont {Engel}, \citenamefont {McLaughlin}, \citenamefont {Mumpower}, \citenamefont {Ney},\ and\ \citenamefont {Surman}}]{Lund2023_InfluenceBetaDecay}%
  \BibitemOpen
  \bibfield  {author} {\bibinfo {author} {\bibfnamefont {K.~A.}\ \bibnamefont {Lund}}, \bibinfo {author} {\bibfnamefont {J.}~\bibnamefont {Engel}}, \bibinfo {author} {\bibfnamefont {G.~C.}\ \bibnamefont {McLaughlin}}, \bibinfo {author} {\bibfnamefont {M.~R.}\ \bibnamefont {Mumpower}}, \bibinfo {author} {\bibfnamefont {E.~M.}\ \bibnamefont {Ney}},\ and\ \bibinfo {author} {\bibfnamefont {R.}~\bibnamefont {Surman}},\ }\bibfield  {title} {\bibinfo {title} {The {{Influence}} of {{Beta Decay Rates}} on r-{{Process Observables}}},\ }\href {https://doi.org/10.3847/1538-4357/acaf56} {\bibfield  {journal} {\bibinfo  {journal} {Astrophys. J.}\ }\textbf {\bibinfo {volume} {944}},\ \bibinfo {pages} {144} (\bibinfo {year} {2023})}\BibitemShut {NoStop}%
\bibitem [{\citenamefont {Goriely}(1997)}]{Goriely1997_DirectNeutronCaptures}%
  \BibitemOpen
  \bibfield  {author} {\bibinfo {author} {\bibfnamefont {S.}~\bibnamefont {Goriely}},\ }\bibfield  {title} {\bibinfo {title} {Direct neutron captures and the r-process nucleosynthesis},\ }\href {https://ui.adsabs.harvard.edu/abs/1997A&A...325..414G} {\bibfield  {journal} {\bibinfo  {journal} {Astron. \& Astrophys.}\ }\textbf {\bibinfo {volume} {325}},\ \bibinfo {pages} {414} (\bibinfo {year} {1997})}\BibitemShut {NoStop}%
\bibitem [{\citenamefont {Surman}\ \emph {et~al.}(1997)\citenamefont {Surman}, \citenamefont {Engel}, \citenamefont {Bennett},\ and\ \citenamefont {Meyer}}]{Surman1997_SourceRareEarthElement}%
  \BibitemOpen
  \bibfield  {author} {\bibinfo {author} {\bibfnamefont {R.}~\bibnamefont {Surman}}, \bibinfo {author} {\bibfnamefont {J.}~\bibnamefont {Engel}}, \bibinfo {author} {\bibfnamefont {J.~R.}\ \bibnamefont {Bennett}},\ and\ \bibinfo {author} {\bibfnamefont {B.~S.}\ \bibnamefont {Meyer}},\ }\bibfield  {title} {\bibinfo {title} {Source of the {{Rare-Earth Element Peak}} in $r$-{{Process Nucleosynthesis}}},\ }\href {https://doi.org/10.1103/PhysRevLett.79.1809} {\bibfield  {journal} {\bibinfo  {journal} {Phys. Rev. Lett.}\ }\textbf {\bibinfo {volume} {79}},\ \bibinfo {pages} {1809} (\bibinfo {year} {1997})}\BibitemShut {NoStop}%
\bibitem [{\citenamefont {Goriely}(1998)}]{Goriely:1998utv}%
  \BibitemOpen
  \bibfield  {author} {\bibinfo {author} {\bibfnamefont {S.}~\bibnamefont {Goriely}},\ }\bibfield  {title} {\bibinfo {title} {{Radiative neutron captures by neutron-rich nuclei and the r-process nucleosynthesis}},\ }\href {https://doi.org/10.1016/S0370-2693(98)00907-1} {\bibfield  {journal} {\bibinfo  {journal} {Phys. Lett. B}\ }\textbf {\bibinfo {volume} {436}},\ \bibinfo {pages} {10} (\bibinfo {year} {1998})}\BibitemShut {NoStop}%
\bibitem [{\citenamefont {Mumpower}\ \emph {et~al.}(2012)\citenamefont {Mumpower}, \citenamefont {McLaughlin},\ and\ \citenamefont {Surman}}]{Mumpower2012_InfluenceNeutronCapture}%
  \BibitemOpen
  \bibfield  {author} {\bibinfo {author} {\bibfnamefont {M.~R.}\ \bibnamefont {Mumpower}}, \bibinfo {author} {\bibfnamefont {G.~C.}\ \bibnamefont {McLaughlin}},\ and\ \bibinfo {author} {\bibfnamefont {R.}~\bibnamefont {Surman}},\ }\bibfield  {title} {\bibinfo {title} {Influence of neutron capture rates in the rare earth region on the r-process abundance pattern},\ }\href {https://doi.org/10.1103/PhysRevC.86.035803} {\bibfield  {journal} {\bibinfo  {journal} {Phys. Rev. C}\ }\textbf {\bibinfo {volume} {86}},\ \bibinfo {pages} {035803} (\bibinfo {year} {2012})}\BibitemShut {NoStop}%
\bibitem [{\citenamefont {Surman}\ \emph {et~al.}(2014)\citenamefont {Surman}, \citenamefont {Mumpower}, \citenamefont {Sinclair}, \citenamefont {Jones}, \citenamefont {Hix},\ and\ \citenamefont {McLaughlin}}]{Surman2014_SensitivityStudiesWeak}%
  \BibitemOpen
  \bibfield  {author} {\bibinfo {author} {\bibfnamefont {R.}~\bibnamefont {Surman}}, \bibinfo {author} {\bibfnamefont {M.}~\bibnamefont {Mumpower}}, \bibinfo {author} {\bibfnamefont {R.}~\bibnamefont {Sinclair}}, \bibinfo {author} {\bibfnamefont {K.~L.}\ \bibnamefont {Jones}}, \bibinfo {author} {\bibfnamefont {W.~R.}\ \bibnamefont {Hix}},\ and\ \bibinfo {author} {\bibfnamefont {G.~C.}\ \bibnamefont {McLaughlin}},\ }\bibfield  {title} {\bibinfo {title} {Sensitivity studies for the weak r-process: Neutron capture rates},\ }\href {https://doi.org/10.1063/1.4867191} {\bibfield  {journal} {\bibinfo  {journal} {AIP Advances}\ }\textbf {\bibinfo {volume} {4}},\ \bibinfo {pages} {041008} (\bibinfo {year} {2014})}\BibitemShut {NoStop}%
\bibitem [{\citenamefont {Nikas}\ \emph {et~al.}(2020)\citenamefont {Nikas}, \citenamefont {Perdikakis}, \citenamefont {Beard}, \citenamefont {Surman}, \citenamefont {Mumpower},\ and\ \citenamefont {Tsintari}}]{Nikas:2020zcc}%
  \BibitemOpen
  \bibfield  {author} {\bibinfo {author} {\bibfnamefont {S.}~\bibnamefont {Nikas}}, \bibinfo {author} {\bibfnamefont {G.}~\bibnamefont {Perdikakis}}, \bibinfo {author} {\bibfnamefont {M.}~\bibnamefont {Beard}}, \bibinfo {author} {\bibfnamefont {R.}~\bibnamefont {Surman}}, \bibinfo {author} {\bibfnamefont {M.~R.}\ \bibnamefont {Mumpower}},\ and\ \bibinfo {author} {\bibfnamefont {P.}~\bibnamefont {Tsintari}},\ }\href@noop {} {\bibinfo {title} {{Propagation of Hauser-Feshbach uncertainty estimates to r-process nucleosynthesis: Benchmark of statistical property models for neutron rich nuclei far from stability}}} (\bibinfo {year} {2020}),\ \Eprint {https://arxiv.org/abs/2010.01698} {arXiv:2010.01698} \BibitemShut {NoStop}%
\bibitem [{\citenamefont {Wiedeking}\ and\ \citenamefont {Goriely}(2024)}]{Wiedeking:2024bty}%
  \BibitemOpen
  \bibfield  {author} {\bibinfo {author} {\bibfnamefont {M.}~\bibnamefont {Wiedeking}}\ and\ \bibinfo {author} {\bibfnamefont {S.}~\bibnamefont {Goriely}},\ }\bibfield  {title} {\bibinfo {title} {{Photon strength functions and nuclear level densities: invaluable input for nucleosynthesis}},\ }\href {https://doi.org/10.1098/rsta.2023.0125} {\bibfield  {journal} {\bibinfo  {journal} {Phil. Trans. Roy. Soc. Lond. A}\ }\textbf {\bibinfo {volume} {382}},\ \bibinfo {pages} {20230125} (\bibinfo {year} {2024})}\BibitemShut {NoStop}%
\bibitem [{\citenamefont {Goriely}\ \emph {et~al.}(2013)\citenamefont {Goriely}, \citenamefont {Sida}, \citenamefont {Lema{\^i}tre}, \citenamefont {Panebianco}, \citenamefont {Dubray}, \citenamefont {Hilaire}, \citenamefont {Bauswein},\ and\ \citenamefont {Janka}}]{Goriely2013_NewFissionFragment}%
  \BibitemOpen
  \bibfield  {author} {\bibinfo {author} {\bibfnamefont {S.}~\bibnamefont {Goriely}}, \bibinfo {author} {\bibfnamefont {J.-L.}\ \bibnamefont {Sida}}, \bibinfo {author} {\bibfnamefont {J.-F.}\ \bibnamefont {Lema{\^i}tre}}, \bibinfo {author} {\bibfnamefont {S.}~\bibnamefont {Panebianco}}, \bibinfo {author} {\bibfnamefont {N.}~\bibnamefont {Dubray}}, \bibinfo {author} {\bibfnamefont {S.}~\bibnamefont {Hilaire}}, \bibinfo {author} {\bibfnamefont {A.}~\bibnamefont {Bauswein}},\ and\ \bibinfo {author} {\bibfnamefont {H.-T.}\ \bibnamefont {Janka}},\ }\bibfield  {title} {\bibinfo {title} {New {{Fission Fragment Distributions}} and $r$-{{Process Origin}} of the {{Rare-Earth Elements}}},\ }\href {https://doi.org/10.1103/PhysRevLett.111.242502} {\bibfield  {journal} {\bibinfo  {journal} {Phys. Rev. Lett.}\ }\textbf {\bibinfo {volume} {111}},\ \bibinfo {pages} {242502} (\bibinfo {year} {2013})}\BibitemShut {NoStop}%
\bibitem [{\citenamefont {Eichler}\ \emph {et~al.}(2015)\citenamefont {Eichler}, \citenamefont {Arcones}, \citenamefont {Kelic}, \citenamefont {Korobkin}, \citenamefont {Langanke}, \citenamefont {Marketin}, \citenamefont {{Martinez-Pinedo}}, \citenamefont {Panov}, \citenamefont {Rauscher}, \citenamefont {Rosswog}, \citenamefont {Winteler}, \citenamefont {Zinner},\ and\ \citenamefont {Thielemann}}]{Eichler2015_RoleFissionNeutron}%
  \BibitemOpen
  \bibfield  {author} {\bibinfo {author} {\bibfnamefont {M.}~\bibnamefont {Eichler}}, \bibinfo {author} {\bibfnamefont {A.}~\bibnamefont {Arcones}}, \bibinfo {author} {\bibfnamefont {A.}~\bibnamefont {Kelic}}, \bibinfo {author} {\bibfnamefont {O.}~\bibnamefont {Korobkin}}, \bibinfo {author} {\bibfnamefont {K.}~\bibnamefont {Langanke}}, \bibinfo {author} {\bibfnamefont {T.}~\bibnamefont {Marketin}}, \bibinfo {author} {\bibfnamefont {G.}~\bibnamefont {{Martinez-Pinedo}}}, \bibinfo {author} {\bibfnamefont {I.}~\bibnamefont {Panov}}, \bibinfo {author} {\bibfnamefont {T.}~\bibnamefont {Rauscher}}, \bibinfo {author} {\bibfnamefont {S.}~\bibnamefont {Rosswog}}, \bibinfo {author} {\bibfnamefont {C.}~\bibnamefont {Winteler}}, \bibinfo {author} {\bibfnamefont {N.~T.}\ \bibnamefont {Zinner}},\ and\ \bibinfo {author} {\bibfnamefont {F.-K.}\ \bibnamefont {Thielemann}},\ }\bibfield  {title} {\bibinfo {title} {The {{Role}} of {{Fission}} in {{Neutron Star Mergers}} and {{Its Impact}} on the {{r}}-{{Process Peaks}}},\ }\href
  {https://doi.org/10.1088/0004-637X/808/1/30} {\bibfield  {journal} {\bibinfo  {journal} {Astrophys. J.}\ }\textbf {\bibinfo {volume} {808}},\ \bibinfo {pages} {30} (\bibinfo {year} {2015})}\BibitemShut {NoStop}%
\bibitem [{\citenamefont {Vassh}\ \emph {et~al.}(2019)\citenamefont {Vassh}, \citenamefont {Vogt}, \citenamefont {Surman}, \citenamefont {Randrup}, \citenamefont {Sprouse}, \citenamefont {Mumpower}, \citenamefont {Jaffke}, \citenamefont {Shaw}, \citenamefont {Holmbeck}, \citenamefont {Zhu},\ and\ \citenamefont {McLaughlin}}]{Vassh2019_UsingExcitationenergyDependent}%
  \BibitemOpen
  \bibfield  {author} {\bibinfo {author} {\bibfnamefont {N.}~\bibnamefont {Vassh}}, \bibinfo {author} {\bibfnamefont {R.}~\bibnamefont {Vogt}}, \bibinfo {author} {\bibfnamefont {R.}~\bibnamefont {Surman}}, \bibinfo {author} {\bibfnamefont {J.}~\bibnamefont {Randrup}}, \bibinfo {author} {\bibfnamefont {T.~M.}\ \bibnamefont {Sprouse}}, \bibinfo {author} {\bibfnamefont {M.~R.}\ \bibnamefont {Mumpower}}, \bibinfo {author} {\bibfnamefont {P.}~\bibnamefont {Jaffke}}, \bibinfo {author} {\bibfnamefont {D.}~\bibnamefont {Shaw}}, \bibinfo {author} {\bibfnamefont {E.~M.}\ \bibnamefont {Holmbeck}}, \bibinfo {author} {\bibfnamefont {Y.}~\bibnamefont {Zhu}},\ and\ \bibinfo {author} {\bibfnamefont {G.~C.}\ \bibnamefont {McLaughlin}},\ }\bibfield  {title} {\bibinfo {title} {Using excitation-energy dependent fission yields to identify key fissioning nuclei in $r$-process nucleosynthesis},\ }\href {https://doi.org/10.1088/1361-6471/ab0bea} {\bibfield  {journal} {\bibinfo  {journal} {J. Phys. G}\ }\textbf {\bibinfo {volume}
  {46}},\ \bibinfo {pages} {065202} (\bibinfo {year} {2019})}\BibitemShut {NoStop}%
\bibitem [{\citenamefont {Giuliani}\ \emph {et~al.}(2020)\citenamefont {Giuliani}, \citenamefont {{Mart{\'i}nez-Pinedo}}, \citenamefont {Wu},\ and\ \citenamefont {Robledo}}]{Giuliani2020_Fission$r$processNucleosynthesis}%
  \BibitemOpen
  \bibfield  {author} {\bibinfo {author} {\bibfnamefont {S.~A.}\ \bibnamefont {Giuliani}}, \bibinfo {author} {\bibfnamefont {G.}~\bibnamefont {{Mart{\'i}nez-Pinedo}}}, \bibinfo {author} {\bibfnamefont {M.-R.}\ \bibnamefont {Wu}},\ and\ \bibinfo {author} {\bibfnamefont {L.~M.}\ \bibnamefont {Robledo}},\ }\bibfield  {title} {\bibinfo {title} {Fission and the $r$-process nucleosynthesis of translead nuclei in neutron star mergers},\ }\href {https://doi.org/10.1103/PhysRevC.102.045804} {\bibfield  {journal} {\bibinfo  {journal} {Phys. Rev. C}\ }\textbf {\bibinfo {volume} {102}},\ \bibinfo {pages} {045804} (\bibinfo {year} {2020})}\BibitemShut {NoStop}%
\bibitem [{\citenamefont {Holmbeck}\ \emph {et~al.}(2023)\citenamefont {Holmbeck}, \citenamefont {Barnes}, \citenamefont {Lund}, \citenamefont {Sprouse}, \citenamefont {McLaughlin},\ and\ \citenamefont {Mumpower}}]{Holmbeck2023_SuperheavyElementsKilonovae}%
  \BibitemOpen
  \bibfield  {author} {\bibinfo {author} {\bibfnamefont {E.~M.}\ \bibnamefont {Holmbeck}}, \bibinfo {author} {\bibfnamefont {J.}~\bibnamefont {Barnes}}, \bibinfo {author} {\bibfnamefont {K.~A.}\ \bibnamefont {Lund}}, \bibinfo {author} {\bibfnamefont {T.~M.}\ \bibnamefont {Sprouse}}, \bibinfo {author} {\bibfnamefont {G.~C.}\ \bibnamefont {McLaughlin}},\ and\ \bibinfo {author} {\bibfnamefont {M.~R.}\ \bibnamefont {Mumpower}},\ }\bibfield  {title} {\bibinfo {title} {Superheavy {{Elements}} in {{Kilonovae}}},\ }\href {https://doi.org/10.3847/2041-8213/acd9cb} {\bibfield  {journal} {\bibinfo  {journal} {Astrophys. J. Lett.}\ }\textbf {\bibinfo {volume} {951}},\ \bibinfo {pages} {L13} (\bibinfo {year} {2023})}\BibitemShut {NoStop}%
\bibitem [{\citenamefont {Xu}\ \emph {et~al.}(2014)\citenamefont {Xu} \emph {et~al.}}]{Xu:2014hla}%
  \BibitemOpen
  \bibfield  {author} {\bibinfo {author} {\bibfnamefont {Z.~Y.}\ \bibnamefont {Xu}} \emph {et~al.},\ }\bibfield  {title} {\bibinfo {title} {{{\ensuremath{\beta}}-Decay Half-Lives of $^{76,77}$Co, $^{79,80}$Ni, and $^{81}$Cu: Experimental Indication of a Doubly Magic $^{78}$Ni}},\ }\href {https://doi.org/10.1103/PhysRevLett.113.032505} {\bibfield  {journal} {\bibinfo  {journal} {Phys. Rev. Lett.}\ }\textbf {\bibinfo {volume} {113}},\ \bibinfo {pages} {032505} (\bibinfo {year} {2014})}\BibitemShut {NoStop}%
\bibitem [{\citenamefont {Atanasov}\ \emph {et~al.}(2015)\citenamefont {Atanasov} \emph {et~al.}}]{Atanasov:2015eza}%
  \BibitemOpen
  \bibfield  {author} {\bibinfo {author} {\bibfnamefont {D.}~\bibnamefont {Atanasov}} \emph {et~al.},\ }\bibfield  {title} {\bibinfo {title} {{Precision Mass Measurements of $^{129-131}$Cd and Their Impact on Stellar Nucleosynthesis via the Rapid Neutron Capture Process}},\ }\href {https://doi.org/10.1103/PhysRevLett.115.232501} {\bibfield  {journal} {\bibinfo  {journal} {Phys. Rev. Lett.}\ }\textbf {\bibinfo {volume} {115}},\ \bibinfo {pages} {232501} (\bibinfo {year} {2015})}\BibitemShut {NoStop}%
\bibitem [{\citenamefont {Lorusso}\ \emph {et~al.}(2015)\citenamefont {Lorusso} \emph {et~al.}}]{Lorusso:2015cna}%
  \BibitemOpen
  \bibfield  {author} {\bibinfo {author} {\bibfnamefont {G.}~\bibnamefont {Lorusso}} \emph {et~al.},\ }\bibfield  {title} {\bibinfo {title} {{{\ensuremath{\beta}}-Decay Half-Lives of 110 Neutron-Rich Nuclei across the $N=82$ Shell Gap: Implications for the Mechanism and Universality of the Astrophysical r-Process}},\ }\href {https://doi.org/10.1103/PhysRevLett.114.192501} {\bibfield  {journal} {\bibinfo  {journal} {Phys. Rev. Lett.}\ }\textbf {\bibinfo {volume} {114}},\ \bibinfo {pages} {192501} (\bibinfo {year} {2015})}\BibitemShut {NoStop}%
\bibitem [{\citenamefont {Welker}\ \emph {et~al.}(2017)\citenamefont {Welker} \emph {et~al.}}]{Welker:2017eja}%
  \BibitemOpen
  \bibfield  {author} {\bibinfo {author} {\bibfnamefont {A.}~\bibnamefont {Welker}} \emph {et~al.},\ }\bibfield  {title} {\bibinfo {title} {{Binding Energy of $^{79}\mathrm{Cu}$: Probing the Structure of the Doubly Magic $^{78}\mathrm{Ni}$ from Only One Proton Away}},\ }\href {https://doi.org/10.1103/PhysRevLett.119.192502} {\bibfield  {journal} {\bibinfo  {journal} {Phys. Rev. Lett.}\ }\textbf {\bibinfo {volume} {119}},\ \bibinfo {pages} {192502} (\bibinfo {year} {2017})}\BibitemShut {NoStop}%
\bibitem [{\citenamefont {Schiller}\ \emph {et~al.}(2000)\citenamefont {Schiller}, \citenamefont {Bergholt}, \citenamefont {Guttormsen}, \citenamefont {Melby}, \citenamefont {Rekstad},\ and\ \citenamefont {Siem}}]{Schiller2000_ExtractionLevelDensity}%
  \BibitemOpen
  \bibfield  {author} {\bibinfo {author} {\bibfnamefont {A.}~\bibnamefont {Schiller}}, \bibinfo {author} {\bibfnamefont {L.}~\bibnamefont {Bergholt}}, \bibinfo {author} {\bibfnamefont {M.}~\bibnamefont {Guttormsen}}, \bibinfo {author} {\bibfnamefont {E.}~\bibnamefont {Melby}}, \bibinfo {author} {\bibfnamefont {J.}~\bibnamefont {Rekstad}},\ and\ \bibinfo {author} {\bibfnamefont {S.}~\bibnamefont {Siem}},\ }\bibfield  {title} {\bibinfo {title} {Extraction of level density and {$\gamma$} strength function from primary {$\gamma$} spectra},\ }\href {https://doi.org/10.1016/S0168-9002(99)01187-0} {\bibfield  {journal} {\bibinfo  {journal} {Nucl. Instrum. \& Methods in Phys. Research Sect. A}\ }\textbf {\bibinfo {volume} {447}},\ \bibinfo {pages} {498} (\bibinfo {year} {2000})}\BibitemShut {NoStop}%
\bibitem [{\citenamefont {Spyrou}\ \emph {et~al.}(2014)\citenamefont {Spyrou}, \citenamefont {Liddick}, \citenamefont {Larsen}, \citenamefont {Guttormsen}, \citenamefont {Cooper}, \citenamefont {Dombos}, \citenamefont {Morrissey}, \citenamefont {Naqvi}, \citenamefont {Perdikakis}, \citenamefont {Quinn}, \citenamefont {Renstr{\o}m}, \citenamefont {Rodriguez}, \citenamefont {Simon}, \citenamefont {Sumithrarachchi},\ and\ \citenamefont {Zegers}}]{Spyrou2014_NovelTechniqueConstraining}%
  \BibitemOpen
  \bibfield  {author} {\bibinfo {author} {\bibfnamefont {A.}~\bibnamefont {Spyrou}}, \bibinfo {author} {\bibfnamefont {S.~N.}\ \bibnamefont {Liddick}}, \bibinfo {author} {\bibfnamefont {A.~C.}\ \bibnamefont {Larsen}}, \bibinfo {author} {\bibfnamefont {M.}~\bibnamefont {Guttormsen}}, \bibinfo {author} {\bibfnamefont {K.}~\bibnamefont {Cooper}}, \bibinfo {author} {\bibfnamefont {A.~C.}\ \bibnamefont {Dombos}}, \bibinfo {author} {\bibfnamefont {D.~J.}\ \bibnamefont {Morrissey}}, \bibinfo {author} {\bibfnamefont {F.}~\bibnamefont {Naqvi}}, \bibinfo {author} {\bibfnamefont {G.}~\bibnamefont {Perdikakis}}, \bibinfo {author} {\bibfnamefont {S.~J.}\ \bibnamefont {Quinn}}, \bibinfo {author} {\bibfnamefont {T.}~\bibnamefont {Renstr{\o}m}}, \bibinfo {author} {\bibfnamefont {J.~A.}\ \bibnamefont {Rodriguez}}, \bibinfo {author} {\bibfnamefont {A.}~\bibnamefont {Simon}}, \bibinfo {author} {\bibfnamefont {C.~S.}\ \bibnamefont {Sumithrarachchi}},\ and\ \bibinfo {author} {\bibfnamefont {R.~G.~T.}\ \bibnamefont {Zegers}},\
  }\bibfield  {title} {\bibinfo {title} {Novel technique for {{Constraining}} $r$-{{Process}} ($n$, $\gamma$) {{Reaction Rates}}},\ }\href {https://doi.org/10.1103/PhysRevLett.113.232502} {\bibfield  {journal} {\bibinfo  {journal} {Phys. Rev. Lett.}\ }\textbf {\bibinfo {volume} {113}},\ \bibinfo {pages} {232502} (\bibinfo {year} {2014})}\BibitemShut {NoStop}%
\bibitem [{\citenamefont {Larsen}\ \emph {et~al.}(2019)\citenamefont {Larsen}, \citenamefont {Spyrou}, \citenamefont {Liddick},\ and\ \citenamefont {Guttormsen}}]{Larsen2019_NovelTechniquesConstraining}%
  \BibitemOpen
  \bibfield  {author} {\bibinfo {author} {\bibfnamefont {A.~C.}\ \bibnamefont {Larsen}}, \bibinfo {author} {\bibfnamefont {A.}~\bibnamefont {Spyrou}}, \bibinfo {author} {\bibfnamefont {S.~N.}\ \bibnamefont {Liddick}},\ and\ \bibinfo {author} {\bibfnamefont {M.}~\bibnamefont {Guttormsen}},\ }\bibfield  {title} {\bibinfo {title} {Novel techniques for constraining neutron-capture rates relevant for $r$-process heavy-element nucleosynthesis},\ }\href {https://doi.org/10.1016/j.ppnp.2019.04.002} {\bibfield  {journal} {\bibinfo  {journal} {Progress in Particle and Nucl. Phys.}\ }\textbf {\bibinfo {volume} {107}},\ \bibinfo {pages} {69} (\bibinfo {year} {2019})}\BibitemShut {NoStop}%
\bibitem [{\citenamefont {Greaves}\ \emph {et~al.}(2025)\citenamefont {Greaves} \emph {et~al.}}]{Greaves2025_ProbingLimitsStatistical}%
  \BibitemOpen
  \bibfield  {author} {\bibinfo {author} {\bibfnamefont {B.}~\bibnamefont {Greaves}} \emph {et~al.},\ }\bibfield  {title} {\bibinfo {title} {Probing the limits of statistical neutron capture for the $r$-process: {{Experimental}} constraints on $^{141}${Cs} nuclear level densities},\ }\href {https://doi.org/10.1016/j.physletb.2025.139992} {\bibfield  {journal} {\bibinfo  {journal} {Physics Lett. B}\ }\textbf {\bibinfo {volume} {871}},\ \bibinfo {pages} {139992} (\bibinfo {year} {2025})}\BibitemShut {NoStop}%
\bibitem [{\citenamefont {Hauser}\ and\ \citenamefont {Feshbach}(1952)}]{Hauser:1952zz}%
  \BibitemOpen
  \bibfield  {author} {\bibinfo {author} {\bibfnamefont {W.}~\bibnamefont {Hauser}}\ and\ \bibinfo {author} {\bibfnamefont {H.}~\bibnamefont {Feshbach}},\ }\bibfield  {title} {\bibinfo {title} {{The Inelastic Scattering of Neutrons}},\ }\href {https://doi.org/10.1103/PhysRev.87.366} {\bibfield  {journal} {\bibinfo  {journal} {Phys. Rev.}\ }\textbf {\bibinfo {volume} {87}},\ \bibinfo {pages} {366} (\bibinfo {year} {1952})}\BibitemShut {NoStop}%
\bibitem [{\citenamefont {Escher}\ \emph {et~al.}(2018)\citenamefont {Escher}, \citenamefont {Burke}, \citenamefont {Hughes}, \citenamefont {Scielzo}, \citenamefont {Casperson}, \citenamefont {Ota}, \citenamefont {Park}, \citenamefont {Saastamoinen},\ and\ \citenamefont {Ross}}]{Escher:2018zhz}%
  \BibitemOpen
  \bibfield  {author} {\bibinfo {author} {\bibfnamefont {J.~E.}\ \bibnamefont {Escher}}, \bibinfo {author} {\bibfnamefont {J.~T.}\ \bibnamefont {Burke}}, \bibinfo {author} {\bibfnamefont {R.~O.}\ \bibnamefont {Hughes}}, \bibinfo {author} {\bibfnamefont {N.~D.}\ \bibnamefont {Scielzo}}, \bibinfo {author} {\bibfnamefont {R.~J.}\ \bibnamefont {Casperson}}, \bibinfo {author} {\bibfnamefont {S.}~\bibnamefont {Ota}}, \bibinfo {author} {\bibfnamefont {H.~I.}\ \bibnamefont {Park}}, \bibinfo {author} {\bibfnamefont {A.}~\bibnamefont {Saastamoinen}},\ and\ \bibinfo {author} {\bibfnamefont {T.~J.}\ \bibnamefont {Ross}},\ }\bibfield  {title} {\bibinfo {title} {{Constraining Neutron Capture Cross Sections for Unstable Nuclei with Surrogate Reaction Data and Theory}},\ }\href {https://doi.org/10.1103/PhysRevLett.121.052501} {\bibfield  {journal} {\bibinfo  {journal} {Phys. Rev. Lett.}\ }\textbf {\bibinfo {volume} {121}},\ \bibinfo {pages} {052501} (\bibinfo {year} {2018})}\BibitemShut {NoStop}%
\bibitem [{\citenamefont {Ratkiewicz}\ \emph {et~al.}(2019)\citenamefont {Ratkiewicz} \emph {et~al.}}]{Ratkiewicz:2018wpx}%
  \BibitemOpen
  \bibfield  {author} {\bibinfo {author} {\bibfnamefont {A.}~\bibnamefont {Ratkiewicz}} \emph {et~al.},\ }\bibfield  {title} {\bibinfo {title} {{Towards Neutron Capture on Exotic Nuclei: Demonstrating $(d,p\gamma)$ as a Surrogate Reaction for $(n,\gamma)$}},\ }\href {https://doi.org/10.1103/PhysRevLett.122.052502} {\bibfield  {journal} {\bibinfo  {journal} {Phys. Rev. Lett.}\ }\textbf {\bibinfo {volume} {122}},\ \bibinfo {pages} {052502} (\bibinfo {year} {2019})}\BibitemShut {NoStop}%
\bibitem [{\citenamefont {Zhang}\ \emph {et~al.}(2025)\citenamefont {Zhang}, \citenamefont {Luo}, \citenamefont {Pang}, \citenamefont {Li}, \citenamefont {Feng}, \citenamefont {JIang}, \citenamefont {Li}, \citenamefont {Xu},\ and\ \citenamefont {Sun}}]{Zhang:2025jta}%
  \BibitemOpen
  \bibfield  {author} {\bibinfo {author} {\bibfnamefont {S.-T.}\ \bibnamefont {Zhang}}, \bibinfo {author} {\bibfnamefont {W.}~\bibnamefont {Luo}}, \bibinfo {author} {\bibfnamefont {D.-Y.}\ \bibnamefont {Pang}}, \bibinfo {author} {\bibfnamefont {Z.-C.}\ \bibnamefont {Li}}, \bibinfo {author} {\bibfnamefont {J.}~\bibnamefont {Feng}}, \bibinfo {author} {\bibfnamefont {B.}~\bibnamefont {JIang}}, \bibinfo {author} {\bibfnamefont {X.-X.}\ \bibnamefont {Li}}, \bibinfo {author} {\bibfnamefont {Y.}~\bibnamefont {Xu}},\ and\ \bibinfo {author} {\bibfnamefont {B.-H.}\ \bibnamefont {Sun}},\ }\href@noop {} {\bibinfo {title} {{Constraining Neutron Capture Cross Sections for $^{88}\mathrm{Y}$ with Gamma-ray Strength Function in $(p,p^\prime\gamma)$ Surrogate Reaction}}} (\bibinfo {year} {2025}),\ \Eprint {https://arxiv.org/abs/2511.10955} {arXiv:2511.10955} \BibitemShut {NoStop}%
\bibitem [{\citenamefont {Baur}(1986)}]{Baur1986_BreakupReactionsIndirect}%
  \BibitemOpen
  \bibfield  {author} {\bibinfo {author} {\bibfnamefont {G.}~\bibnamefont {Baur}},\ }\bibfield  {title} {\bibinfo {title} {Breakup reactions as an indirect method to investigate low-energy charged-particle reactions relevant for nuclear astrophysics},\ }\href {https://doi.org/10.1016/0370-2693(86)91483-8} {\bibfield  {journal} {\bibinfo  {journal} {Physics Lett. B}\ }\textbf {\bibinfo {volume} {178}},\ \bibinfo {pages} {135} (\bibinfo {year} {1986})}\BibitemShut {NoStop}%
\bibitem [{\citenamefont {Xu}\ \emph {et~al.}(1994)\citenamefont {Xu}, \citenamefont {Gagliardi}, \citenamefont {Tribble}, \citenamefont {Mukhamedzhanov},\ and\ \citenamefont {Timofeyuk}}]{Xu1994_OverallNormalizationAstrophysical}%
  \BibitemOpen
  \bibfield  {author} {\bibinfo {author} {\bibfnamefont {H.~M.}\ \bibnamefont {Xu}}, \bibinfo {author} {\bibfnamefont {C.~A.}\ \bibnamefont {Gagliardi}}, \bibinfo {author} {\bibfnamefont {R.~E.}\ \bibnamefont {Tribble}}, \bibinfo {author} {\bibfnamefont {A.~M.}\ \bibnamefont {Mukhamedzhanov}},\ and\ \bibinfo {author} {\bibfnamefont {N.~K.}\ \bibnamefont {Timofeyuk}},\ }\bibfield  {title} {\bibinfo {title} {Overall {{Normalization}} of the {{Astrophysical S-Factor}} and the {{Nuclear Vertex Constant}} for $^7${Be}(p,{$\gamma$})$^8${B} {Reactions}},\ }\href {https://doi.org/10.1103/PhysRevLett.73.2027} {\bibfield  {journal} {\bibinfo  {journal} {Phys. Rev. Lett.}\ }\textbf {\bibinfo {volume} {73}},\ \bibinfo {pages} {2027} (\bibinfo {year} {1994})}\BibitemShut {NoStop}%
\bibitem [{\citenamefont {Baur}\ \emph {et~al.}(1986)\citenamefont {Baur}, \citenamefont {Bertulani},\ and\ \citenamefont {Rebel}}]{Baur1986_CoulombDissociationSource}%
  \BibitemOpen
  \bibfield  {author} {\bibinfo {author} {\bibfnamefont {G.}~\bibnamefont {Baur}}, \bibinfo {author} {\bibfnamefont {C.~A.}\ \bibnamefont {Bertulani}},\ and\ \bibinfo {author} {\bibfnamefont {H.}~\bibnamefont {Rebel}},\ }\bibfield  {title} {\bibinfo {title} {Coulomb dissociation as a source of information on radiative capture processes of astrophysical interest},\ }\href {https://doi.org/10.1016/0375-9474(86)90290-3} {\bibfield  {journal} {\bibinfo  {journal} {Nuclear Physics A}\ }\textbf {\bibinfo {volume} {458}},\ \bibinfo {pages} {188} (\bibinfo {year} {1986})}\BibitemShut {NoStop}%
\bibitem [{\citenamefont {Reifarth}\ and\ \citenamefont {Litvinov}(2014)}]{Reifarth2014_MeasurementsNeutroninducedReactions}%
  \BibitemOpen
  \bibfield  {author} {\bibinfo {author} {\bibfnamefont {R.}~\bibnamefont {Reifarth}}\ and\ \bibinfo {author} {\bibfnamefont {Y.~A.}\ \bibnamefont {Litvinov}},\ }\bibfield  {title} {\bibinfo {title} {Measurements of neutron-induced reactions in inverse kinematics},\ }\href {https://doi.org/10.1103/PhysRevSTAB.17.014701} {\bibfield  {journal} {\bibinfo  {journal} {Phys. Rev. Spec. Top. - Accel. \& Beams}\ }\textbf {\bibinfo {volume} {17}},\ \bibinfo {pages} {014701} (\bibinfo {year} {2014})}\BibitemShut {NoStop}%
\bibitem [{\citenamefont {Reifarth}\ \emph {et~al.}(2017)\citenamefont {Reifarth}, \citenamefont {G{\"o}bel}, \citenamefont {Heftrich}, \citenamefont {Weigand}, \citenamefont {Jurado}, \citenamefont {K{\"a}ppeler},\ and\ \citenamefont {Litvinov}}]{Reifarth2017_SpallationbasedNeutronTarget}%
  \BibitemOpen
  \bibfield  {author} {\bibinfo {author} {\bibfnamefont {R.}~\bibnamefont {Reifarth}}, \bibinfo {author} {\bibfnamefont {K.}~\bibnamefont {G{\"o}bel}}, \bibinfo {author} {\bibfnamefont {T.}~\bibnamefont {Heftrich}}, \bibinfo {author} {\bibfnamefont {M.}~\bibnamefont {Weigand}}, \bibinfo {author} {\bibfnamefont {B.}~\bibnamefont {Jurado}}, \bibinfo {author} {\bibfnamefont {F.}~\bibnamefont {K{\"a}ppeler}},\ and\ \bibinfo {author} {\bibfnamefont {Y.~A.}\ \bibnamefont {Litvinov}},\ }\bibfield  {title} {\bibinfo {title} {Spallation-based neutron target for direct studies of neutron-induced reactions in inverse kinematics},\ }\href {https://doi.org/10.1103/PhysRevAccelBeams.20.044701} {\bibfield  {journal} {\bibinfo  {journal} {Phys. Rev. Accel. \& Beams}\ }\textbf {\bibinfo {volume} {20}},\ \bibinfo {pages} {044701} (\bibinfo {year} {2017})}\BibitemShut {NoStop}%
\bibitem [{\citenamefont {{Tarife{\~n}o-Saldivia}}\ \emph {et~al.}(2026)\citenamefont {{Tarife{\~n}o-Saldivia}}, \citenamefont {{Domingo-Pardo}}, \citenamefont {Dillmann},\ and\ \citenamefont {Litvinov}}]{Tarifeno-Saldivia2026_DirectNeutronReactions}%
  \BibitemOpen
  \bibfield  {author} {\bibinfo {author} {\bibfnamefont {A.}~\bibnamefont {{Tarife{\~n}o-Saldivia}}}, \bibinfo {author} {\bibfnamefont {C.}~\bibnamefont {{Domingo-Pardo}}}, \bibinfo {author} {\bibfnamefont {I.}~\bibnamefont {Dillmann}},\ and\ \bibinfo {author} {\bibfnamefont {Y.~A.}\ \bibnamefont {Litvinov}},\ }\href@noop {} {\bibinfo {title} {Direct {{Neutron Reactions}} in {{Storage Rings Utilizing}} a {{Supercompact Cyclotron Neutron Target}}}} (\bibinfo {year} {2026}),\ \Eprint {https://arxiv.org/abs/2508.15465} {arXiv:2508.15465} \BibitemShut {NoStop}%
\bibitem [{\citenamefont {Kedia}\ \emph {et~al.}(2026)\citenamefont {Kedia}, \citenamefont {Berryman}, \citenamefont {Cabrera~Garcia}, \citenamefont {Escher}, \citenamefont {Gorton}, \citenamefont {Holmbeck}, \citenamefont {McLaughlin}, \citenamefont {Pruitt}, \citenamefont {Sieverding},\ and\ \citenamefont {Surman}}]{Kedia:2026usu}%
  \BibitemOpen
  \bibfield  {author} {\bibinfo {author} {\bibfnamefont {A.}~\bibnamefont {Kedia}}, \bibinfo {author} {\bibfnamefont {J.~M.}\ \bibnamefont {Berryman}}, \bibinfo {author} {\bibfnamefont {J.}~\bibnamefont {Cabrera~Garcia}}, \bibinfo {author} {\bibfnamefont {J.~E.}\ \bibnamefont {Escher}}, \bibinfo {author} {\bibfnamefont {O.~C.}\ \bibnamefont {Gorton}}, \bibinfo {author} {\bibfnamefont {E.~M.}\ \bibnamefont {Holmbeck}}, \bibinfo {author} {\bibfnamefont {G.~C.}\ \bibnamefont {McLaughlin}}, \bibinfo {author} {\bibfnamefont {C.~D.}\ \bibnamefont {Pruitt}}, \bibinfo {author} {\bibfnamefont {A.}~\bibnamefont {Sieverding}},\ and\ \bibinfo {author} {\bibfnamefont {R.}~\bibnamefont {Surman}},\ }\bibfield  {title} {\bibinfo {title} {Correlated and uncorrelated {{Monte Carlo}} neutron-capture rate variations in weak $r$-process simulations},\ }\href {https://doi.org/10.1103/32jv-8965} {\bibfield  {journal} {\bibinfo  {journal} {Phys. Rev. C}\ }\textbf {\bibinfo {volume} {114}},\ \bibinfo {pages} {035805} (\bibinfo {year}
  {2026})}\BibitemShut {NoStop}%
\bibitem [{\citenamefont {Sieja}\ and\ \citenamefont {Goriely}(2021)}]{Sieja2021_ShellmodelBasedStudy}%
  \BibitemOpen
  \bibfield  {author} {\bibinfo {author} {\bibfnamefont {K.}~\bibnamefont {Sieja}}\ and\ \bibinfo {author} {\bibfnamefont {S.}~\bibnamefont {Goriely}},\ }\bibfield  {title} {\bibinfo {title} {Shell-model based study of the direct capture in neutron-rich nuclei},\ }\href {https://doi.org/10.1140/epja/s10050-021-00439-2} {\bibfield  {journal} {\bibinfo  {journal} {Eur. Phys. J. A}\ }\textbf {\bibinfo {volume} {57}},\ \bibinfo {pages} {110} (\bibinfo {year} {2021})}\BibitemShut {NoStop}%
\bibitem [{\citenamefont {Meyer}(1994)}]{Meyer:1994gn}%
  \BibitemOpen
  \bibfield  {author} {\bibinfo {author} {\bibfnamefont {B.~S.}\ \bibnamefont {Meyer}},\ }\bibfield  {title} {\bibinfo {title} {{The r-, s-, and p-processes in nucleosynthesis}},\ }\href {https://doi.org/10.1146/annurev.aa.32.090194.001101} {\bibfield  {journal} {\bibinfo  {journal} {Ann. Rev. Astron. Astrophys.}\ }\textbf {\bibinfo {volume} {32}},\ \bibinfo {pages} {153} (\bibinfo {year} {1994})}\BibitemShut {NoStop}%
\bibitem [{\citenamefont {Li}\ and\ \citenamefont {Meyer}(2026)}]{Li:2026sur}%
  \BibitemOpen
  \bibfield  {author} {\bibinfo {author} {\bibfnamefont {M.}~\bibnamefont {Li}}\ and\ \bibinfo {author} {\bibfnamefont {B.~S.}\ \bibnamefont {Meyer}},\ }\bibfield  {title} {\bibinfo {title} {{Neutron Economy and Freeze-out Dynamics in the Late-time Cold r-process}},\ }\href {https://doi.org/10.3847/1538-4357/aea157} {\bibfield  {journal} {\bibinfo  {journal} {Astrophys. J.}\ }\textbf {\bibinfo {volume} {1009}},\ \bibinfo {pages} {202} (\bibinfo {year} {2026})}\BibitemShut {NoStop}%
\bibitem [{\citenamefont {Xu}\ \emph {et~al.}(2026)\citenamefont {Xu}, \citenamefont {Du}, \citenamefont {Li},\ and\ \citenamefont {Fang}}]{Xu2026_ImpactNuclearLevel}%
  \BibitemOpen
  \bibfield  {author} {\bibinfo {author} {\bibfnamefont {H.}~\bibnamefont {Xu}}, \bibinfo {author} {\bibfnamefont {P.-X.}\ \bibnamefont {Du}}, \bibinfo {author} {\bibfnamefont {J.}~\bibnamefont {Li}},\ and\ \bibinfo {author} {\bibfnamefont {D.-L.}\ \bibnamefont {Fang}},\ }\bibfield  {title} {\bibinfo {title} {Impact of {{Nuclear Level Density}} on r-process {{Rare}}-earth {{Peak Nucleosynthesis}}},\ }\href {https://doi.org/10.3847/1538-4357/ae9d64} {\bibfield  {journal} {\bibinfo  {journal} {The Astrophys. J.}\ }\textbf {\bibinfo {volume} {1009}},\ \bibinfo {pages} {61} (\bibinfo {year} {2026})}\BibitemShut {NoStop}%
\bibitem [{\citenamefont {Bliss}\ \emph {et~al.}(2020)\citenamefont {Bliss}, \citenamefont {Arcones}, \citenamefont {Montes},\ and\ \citenamefont {Pereira}}]{Bliss2020_NuclearPhysicsUncertainties}%
  \BibitemOpen
  \bibfield  {author} {\bibinfo {author} {\bibfnamefont {J.}~\bibnamefont {Bliss}}, \bibinfo {author} {\bibfnamefont {A.}~\bibnamefont {Arcones}}, \bibinfo {author} {\bibfnamefont {F.}~\bibnamefont {Montes}},\ and\ \bibinfo {author} {\bibfnamefont {J.}~\bibnamefont {Pereira}},\ }\bibfield  {title} {\bibinfo {title} {Nuclear physics uncertainties in neutrino-driven, neutron-rich supernova ejecta},\ }\href {https://doi.org/10.1103/PhysRevC.101.055807} {\bibfield  {journal} {\bibinfo  {journal} {Phys. Rev. C}\ }\textbf {\bibinfo {volume} {101}},\ \bibinfo {pages} {055807} (\bibinfo {year} {2020})}\BibitemShut {NoStop}%
\bibitem [{\citenamefont {Psaltis}\ \emph {et~al.}(2022)\citenamefont {Psaltis}, \citenamefont {Arcones}, \citenamefont {Montes}, \citenamefont {Mohr}, \citenamefont {Hansen}, \citenamefont {Jacobi},\ and\ \citenamefont {Schatz}}]{Psaltis2022_ConstrainingNucleosynthesisNeutrinodriven}%
  \BibitemOpen
  \bibfield  {author} {\bibinfo {author} {\bibfnamefont {A.}~\bibnamefont {Psaltis}}, \bibinfo {author} {\bibfnamefont {A.}~\bibnamefont {Arcones}}, \bibinfo {author} {\bibfnamefont {F.}~\bibnamefont {Montes}}, \bibinfo {author} {\bibfnamefont {P.}~\bibnamefont {Mohr}}, \bibinfo {author} {\bibfnamefont {C.~J.}\ \bibnamefont {Hansen}}, \bibinfo {author} {\bibfnamefont {M.}~\bibnamefont {Jacobi}},\ and\ \bibinfo {author} {\bibfnamefont {H.}~\bibnamefont {Schatz}},\ }\bibfield  {title} {\bibinfo {title} {Constraining {{Nucleosynthesis}} in {{Neutrino}}-driven {{Winds}}: {{Observations}}, {{Simulations}}, and {{Nuclear Physics}}},\ }\href {https://doi.org/10.3847/1538-4357/ac7da7} {\bibfield  {journal} {\bibinfo  {journal} {Astrophys. J.}\ }\textbf {\bibinfo {volume} {935}},\ \bibinfo {pages} {27} (\bibinfo {year} {2022})}\BibitemShut {NoStop}%
\bibitem [{\citenamefont {Reichert}\ \emph {et~al.}(2023)\citenamefont {Reichert} \emph {et~al.}}]{Reichert:2023xqy}%
  \BibitemOpen
  \bibfield  {author} {\bibinfo {author} {\bibfnamefont {M.}~\bibnamefont {Reichert}} \emph {et~al.},\ }\bibfield  {title} {\bibinfo {title} {{The Nuclear Reaction Network WinNet}},\ }\href {https://doi.org/10.3847/1538-4365/acf033} {\bibfield  {journal} {\bibinfo  {journal} {Astrophys. J. Suppl.}\ }\textbf {\bibinfo {volume} {268}},\ \bibinfo {pages} {66} (\bibinfo {year} {2023})}\BibitemShut {NoStop}%
\bibitem [{\citenamefont {Kuske}\ \emph {et~al.}(2025)\citenamefont {Kuske}, \citenamefont {Arcones},\ and\ \citenamefont {Reichert}}]{Kuske:2025ffp}%
  \BibitemOpen
  \bibfield  {author} {\bibinfo {author} {\bibfnamefont {J.}~\bibnamefont {Kuske}}, \bibinfo {author} {\bibfnamefont {A.}~\bibnamefont {Arcones}},\ and\ \bibinfo {author} {\bibfnamefont {M.}~\bibnamefont {Reichert}},\ }\bibfield  {title} {\bibinfo {title} {{Complete Survey of r-process Conditions: The (Un)robustness of the r-process(es)}},\ }\href {https://doi.org/10.3847/1538-4357/adf0f7} {\bibfield  {journal} {\bibinfo  {journal} {Astrophys. J.}\ }\textbf {\bibinfo {volume} {990}},\ \bibinfo {pages} {37} (\bibinfo {year} {2025})}\BibitemShut {NoStop}%
\bibitem [{\citenamefont {Koning}\ \emph {et~al.}(2023)\citenamefont {Koning}, \citenamefont {Hilaire},\ and\ \citenamefont {Goriely}}]{Koning:2023ixl}%
  \BibitemOpen
  \bibfield  {author} {\bibinfo {author} {\bibfnamefont {A.}~\bibnamefont {Koning}}, \bibinfo {author} {\bibfnamefont {S.}~\bibnamefont {Hilaire}},\ and\ \bibinfo {author} {\bibfnamefont {S.}~\bibnamefont {Goriely}},\ }\bibfield  {title} {\bibinfo {title} {{TALYS: modeling of nuclear reactions}},\ }\href {https://doi.org/10.1140/epja/s10050-023-01034-3} {\bibfield  {journal} {\bibinfo  {journal} {Eur. Phys. J. A}\ }\textbf {\bibinfo {volume} {59}},\ \bibinfo {pages} {131} (\bibinfo {year} {2023})}\BibitemShut {NoStop}%
\bibitem [{\citenamefont {Jacobi}\ \emph {et~al.}(2023)\citenamefont {Jacobi}, \citenamefont {Guercilena}, \citenamefont {Huth}, \citenamefont {Ricigliano}, \citenamefont {Arcones},\ and\ \citenamefont {Schwenk}}]{Jacobi:2023olu}%
  \BibitemOpen
  \bibfield  {author} {\bibinfo {author} {\bibfnamefont {M.}~\bibnamefont {Jacobi}}, \bibinfo {author} {\bibfnamefont {F.~M.}\ \bibnamefont {Guercilena}}, \bibinfo {author} {\bibfnamefont {S.}~\bibnamefont {Huth}}, \bibinfo {author} {\bibfnamefont {G.}~\bibnamefont {Ricigliano}}, \bibinfo {author} {\bibfnamefont {A.}~\bibnamefont {Arcones}},\ and\ \bibinfo {author} {\bibfnamefont {A.}~\bibnamefont {Schwenk}},\ }\bibfield  {title} {\bibinfo {title} {{Effects of nuclear matter properties in neutron star mergers}},\ }\href {https://doi.org/10.1093/mnras/stad3738} {\bibfield  {journal} {\bibinfo  {journal} {Mon. Not. Roy. Astron. Soc.}\ }\textbf {\bibinfo {volume} {527}},\ \bibinfo {pages} {8812} (\bibinfo {year} {2023})}\BibitemShut {NoStop}%
\bibitem [{\citenamefont {Ricigliano}\ \emph {et~al.}(2024)\citenamefont {Ricigliano}, \citenamefont {Jacobi},\ and\ \citenamefont {Arcones}}]{Ricigliano:2024lwf}%
  \BibitemOpen
  \bibfield  {author} {\bibinfo {author} {\bibfnamefont {G.}~\bibnamefont {Ricigliano}}, \bibinfo {author} {\bibfnamefont {M.}~\bibnamefont {Jacobi}},\ and\ \bibinfo {author} {\bibfnamefont {A.}~\bibnamefont {Arcones}},\ }\bibfield  {title} {\bibinfo {title} {{Impact of nuclear matter properties on the nucleosynthesis and the kilonova from binary neutron star merger ejecta}},\ }\href {https://doi.org/10.1093/mnras/stae1979} {\bibfield  {journal} {\bibinfo  {journal} {Mon. Not. Roy. Astron. Soc.}\ }\textbf {\bibinfo {volume} {533}},\ \bibinfo {pages} {2096} (\bibinfo {year} {2024})}\BibitemShut {NoStop}%
\bibitem [{\citenamefont {Wu}\ \emph {et~al.}(2016)\citenamefont {Wu}, \citenamefont {Fern{\'a}ndez}, \citenamefont {Mart{\'\i}nez-Pinedo},\ and\ \citenamefont {Metzger}}]{Wu:2016pnw}%
  \BibitemOpen
  \bibfield  {author} {\bibinfo {author} {\bibfnamefont {M.-R.}\ \bibnamefont {Wu}}, \bibinfo {author} {\bibfnamefont {R.}~\bibnamefont {Fern{\'a}ndez}}, \bibinfo {author} {\bibfnamefont {G.}~\bibnamefont {Mart{\'\i}nez-Pinedo}},\ and\ \bibinfo {author} {\bibfnamefont {B.~D.}\ \bibnamefont {Metzger}},\ }\bibfield  {title} {\bibinfo {title} {{Production of the entire range of r-process nuclides by black hole accretion disc outflows from neutron star mergers}},\ }\href {https://doi.org/10.1093/mnras/stw2156} {\bibfield  {journal} {\bibinfo  {journal} {Mon. Not. Roy. Astron. Soc.}\ }\textbf {\bibinfo {volume} {463}},\ \bibinfo {pages} {2323} (\bibinfo {year} {2016})}\BibitemShut {NoStop}%
\bibitem [{\citenamefont {M{\"o}ller}\ \emph {et~al.}(2016)\citenamefont {M{\"o}ller}, \citenamefont {Sierk}, \citenamefont {Ichikawa},\ and\ \citenamefont {Sagawa}}]{Moller:2015fba}%
  \BibitemOpen
  \bibfield  {author} {\bibinfo {author} {\bibfnamefont {P.}~\bibnamefont {M{\"o}ller}}, \bibinfo {author} {\bibfnamefont {A.~J.}\ \bibnamefont {Sierk}}, \bibinfo {author} {\bibfnamefont {T.}~\bibnamefont {Ichikawa}},\ and\ \bibinfo {author} {\bibfnamefont {H.}~\bibnamefont {Sagawa}},\ }\bibfield  {title} {\bibinfo {title} {{Nuclear ground-state masses and deformations: FRDM(2012)}},\ }\href {https://doi.org/10.1016/j.adt.2015.10.002} {\bibfield  {journal} {\bibinfo  {journal} {Atom. Data Nucl. Data Tabl.}\ }\textbf {\bibinfo {volume} {109-110}},\ \bibinfo {pages} {1} (\bibinfo {year} {2016})}\BibitemShut {NoStop}%
\bibitem [{\citenamefont {{Sneden}}\ \emph {et~al.}(2008)\citenamefont {{Sneden}}, \citenamefont {{Cowan}},\ and\ \citenamefont {{Gallino}}}]{2008ARA&A..46..241S}%
  \BibitemOpen
  \bibfield  {author} {\bibinfo {author} {\bibfnamefont {C.}~\bibnamefont {{Sneden}}}, \bibinfo {author} {\bibfnamefont {J.~J.}\ \bibnamefont {{Cowan}}},\ and\ \bibinfo {author} {\bibfnamefont {R.}~\bibnamefont {{Gallino}}},\ }\bibfield  {title} {\bibinfo {title} {{Neutron-capture elements in the early galaxy.}},\ }\href {https://doi.org/10.1146/annurev.astro.46.060407.145207} {\bibfield  {journal} {\bibinfo  {journal} {Ann. Rev. Astron. \& Astrophys.}\ }\textbf {\bibinfo {volume} {46}},\ \bibinfo {pages} {241} (\bibinfo {year} {2008})}\BibitemShut {NoStop}%
\bibitem [{\citenamefont {Goriely}\ \emph {et~al.}(2008)\citenamefont {Goriely}, \citenamefont {Hilaire},\ and\ \citenamefont {Koning}}]{Goriely:2008zz}%
  \BibitemOpen
  \bibfield  {author} {\bibinfo {author} {\bibfnamefont {S.}~\bibnamefont {Goriely}}, \bibinfo {author} {\bibfnamefont {S.}~\bibnamefont {Hilaire}},\ and\ \bibinfo {author} {\bibfnamefont {A.~J.}\ \bibnamefont {Koning}},\ }\bibfield  {title} {\bibinfo {title} {{Improved microscopic nuclear level densities within the Hartree-Fock-Bogoliubov plus combinatorial method}},\ }\href {https://doi.org/10.1103/PhysRevC.78.064307} {\bibfield  {journal} {\bibinfo  {journal} {Phys. Rev. C}\ }\textbf {\bibinfo {volume} {78}},\ \bibinfo {pages} {064307} (\bibinfo {year} {2008})}\BibitemShut {NoStop}%
\bibitem [{\citenamefont {Goriely}\ \emph {et~al.}(2018)\citenamefont {Goriely}, \citenamefont {Hilaire}, \citenamefont {P{\'e}ru},\ and\ \citenamefont {Sieja}}]{Goriely:2018gig}%
  \BibitemOpen
  \bibfield  {author} {\bibinfo {author} {\bibfnamefont {S.}~\bibnamefont {Goriely}}, \bibinfo {author} {\bibfnamefont {S.}~\bibnamefont {Hilaire}}, \bibinfo {author} {\bibfnamefont {S.}~\bibnamefont {P{\'e}ru}},\ and\ \bibinfo {author} {\bibfnamefont {K.}~\bibnamefont {Sieja}},\ }\bibfield  {title} {\bibinfo {title} {{Gogny-HFB+QRPA dipole strength function and its application to radiative nucleon capture cross section}},\ }\href {https://doi.org/10.1103/PhysRevC.98.014327} {\bibfield  {journal} {\bibinfo  {journal} {Phys. Rev. C}\ }\textbf {\bibinfo {volume} {98}},\ \bibinfo {pages} {014327} (\bibinfo {year} {2018})}\BibitemShut {NoStop}%
\bibitem [{\citenamefont {Rochman}\ \emph {et~al.}(2025)\citenamefont {Rochman}, \citenamefont {Koning}, \citenamefont {Goriely},\ and\ \citenamefont {Hilaire}}]{Rochman:2024yua}%
  \BibitemOpen
  \bibfield  {author} {\bibinfo {author} {\bibfnamefont {D.}~\bibnamefont {Rochman}}, \bibinfo {author} {\bibfnamefont {A.}~\bibnamefont {Koning}}, \bibinfo {author} {\bibfnamefont {S.}~\bibnamefont {Goriely}},\ and\ \bibinfo {author} {\bibfnamefont {S.}~\bibnamefont {Hilaire}},\ }\bibfield  {title} {\bibinfo {title} {{TENDL-astro: A new nuclear data set for astrophysics interest}},\ }\href {https://doi.org/10.1016/j.nuclphysa.2024.122951} {\bibfield  {journal} {\bibinfo  {journal} {Nucl. Phys. A}\ }\textbf {\bibinfo {volume} {1053}},\ \bibinfo {pages} {122951} (\bibinfo {year} {2025})}\BibitemShut {NoStop}%
\bibitem [{\citenamefont {Goriely}\ and\ \citenamefont {Plujko}(2019)}]{Goriely:2019upm}%
  \BibitemOpen
  \bibfield  {author} {\bibinfo {author} {\bibfnamefont {S.}~\bibnamefont {Goriely}}\ and\ \bibinfo {author} {\bibfnamefont {V.}~\bibnamefont {Plujko}},\ }\bibfield  {title} {\bibinfo {title} {{Simple empirical E1 and M1 strength functions for practical applications}},\ }\href {https://doi.org/10.1103/PhysRevC.99.014303} {\bibfield  {journal} {\bibinfo  {journal} {Phys. Rev. C}\ }\textbf {\bibinfo {volume} {99}},\ \bibinfo {pages} {014303} (\bibinfo {year} {2019})}\BibitemShut {NoStop}%
\bibitem [{\citenamefont {Goriely}\ \emph {et~al.}(2022)\citenamefont {Goriely}, \citenamefont {Larsen},\ and\ \citenamefont {M{\"u}cher}}]{Goriely:2022flp}%
  \BibitemOpen
  \bibfield  {author} {\bibinfo {author} {\bibfnamefont {S.}~\bibnamefont {Goriely}}, \bibinfo {author} {\bibfnamefont {A.-C.}\ \bibnamefont {Larsen}},\ and\ \bibinfo {author} {\bibfnamefont {D.}~\bibnamefont {M{\"u}cher}},\ }\bibfield  {title} {\bibinfo {title} {{Comprehensive test of nuclear level density models}},\ }\href {https://doi.org/10.1103/PhysRevC.106.044315} {\bibfield  {journal} {\bibinfo  {journal} {Phys. Rev. C}\ }\textbf {\bibinfo {volume} {106}},\ \bibinfo {pages} {044315} (\bibinfo {year} {2022})}\BibitemShut {NoStop}%
\bibitem [{\citenamefont {Markova}\ \emph {et~al.}(2026)\citenamefont {Markova} \emph {et~al.}}]{Markova:2025iwu}%
  \BibitemOpen
  \bibfield  {author} {\bibinfo {author} {\bibfnamefont {M.}~\bibnamefont {Markova}} \emph {et~al.},\ }\bibfield  {title} {\bibinfo {title} {{Exploring the statistical properties of the neutron-deficient $^{109}$In isotope with the Oslo method}},\ }\href {https://doi.org/10.1103/qrvh-s12k} {\bibfield  {journal} {\bibinfo  {journal} {Phys. Rev. C}\ }\textbf {\bibinfo {volume} {113}},\ \bibinfo {pages} {044310} (\bibinfo {year} {2026})}\BibitemShut {NoStop}%
\bibitem [{\citenamefont {Koning}\ \emph {et~al.}(2008)\citenamefont {Koning}, \citenamefont {Hilaire},\ and\ \citenamefont {Goriely}}]{Koning:2008ygi}%
  \BibitemOpen
  \bibfield  {author} {\bibinfo {author} {\bibfnamefont {A.~J.}\ \bibnamefont {Koning}}, \bibinfo {author} {\bibfnamefont {S.}~\bibnamefont {Hilaire}},\ and\ \bibinfo {author} {\bibfnamefont {S.}~\bibnamefont {Goriely}},\ }\bibfield  {title} {\bibinfo {title} {{Global and local level density models}},\ }\href {https://doi.org/10.1016/j.nuclphysa.2008.06.005} {\bibfield  {journal} {\bibinfo  {journal} {Nucl. Phys. A}\ }\textbf {\bibinfo {volume} {810}},\ \bibinfo {pages} {13} (\bibinfo {year} {2008})}\BibitemShut {NoStop}%
\bibitem [{\citenamefont {Demetriou}\ and\ \citenamefont {Goriely}(2001)}]{Demetriou:2001cvr}%
  \BibitemOpen
  \bibfield  {author} {\bibinfo {author} {\bibfnamefont {P.}~\bibnamefont {Demetriou}}\ and\ \bibinfo {author} {\bibfnamefont {S.}~\bibnamefont {Goriely}},\ }\bibfield  {title} {\bibinfo {title} {{Microscopic nuclear level densities for practical applications}},\ }\href {https://doi.org/10.1016/S0375-9474(01)01095-8} {\bibfield  {journal} {\bibinfo  {journal} {Nucl. Phys. A}\ }\textbf {\bibinfo {volume} {695}},\ \bibinfo {pages} {95} (\bibinfo {year} {2001})}\BibitemShut {NoStop}%
\bibitem [{\citenamefont {Ignatyuk}\ \emph {et~al.}(1979)\citenamefont {Ignatyuk}, \citenamefont {Istekov},\ and\ \citenamefont {Smirenkin}}]{osti_5770504}%
  \BibitemOpen
  \bibfield  {author} {\bibinfo {author} {\bibfnamefont {A.~V.}\ \bibnamefont {Ignatyuk}}, \bibinfo {author} {\bibfnamefont {K.~K.}\ \bibnamefont {Istekov}},\ and\ \bibinfo {author} {\bibfnamefont {G.~N.}\ \bibnamefont {Smirenkin}},\ }\bibfield  {title} {\bibinfo {title} {Role of collective effects in the systematics of nuclear level densities},\ }\href {https://inis.iaea.org/records/64xas-6sx16} {\bibfield  {journal} {\bibinfo  {journal} {Sov. J. Nucl. Phys.}\ }\textbf {\bibinfo {volume} {29:4}} (\bibinfo {year} {1979})}\BibitemShut {NoStop}%
\bibitem [{\citenamefont {{Ignatyuk}}\ \emph {et~al.}(1993)\citenamefont {{Ignatyuk}}, \citenamefont {{Weil}}, \citenamefont {{Raman}},\ and\ \citenamefont {{Kahane}}}]{1993PhRvC..47.1504I}%
  \BibitemOpen
  \bibfield  {author} {\bibinfo {author} {\bibfnamefont {A.~V.}\ \bibnamefont {{Ignatyuk}}}, \bibinfo {author} {\bibfnamefont {J.~L.}\ \bibnamefont {{Weil}}}, \bibinfo {author} {\bibfnamefont {S.}~\bibnamefont {{Raman}}},\ and\ \bibinfo {author} {\bibfnamefont {S.}~\bibnamefont {{Kahane}}},\ }\bibfield  {title} {\bibinfo {title} {{Density of discrete levels in $^{116}$Sn}},\ }\href {https://doi.org/10.1103/PhysRevC.47.1504} {\bibfield  {journal} {\bibinfo  {journal} {\prc}\ }\textbf {\bibinfo {volume} {47}},\ \bibinfo {pages} {1504} (\bibinfo {year} {1993})}\BibitemShut {NoStop}%
\bibitem [{\citenamefont {Goriely}\ \emph {et~al.}(2001)\citenamefont {Goriely}, \citenamefont {Tondeur},\ and\ \citenamefont {Pearson}}]{Goriely:2001zz}%
  \BibitemOpen
  \bibfield  {author} {\bibinfo {author} {\bibfnamefont {S.}~\bibnamefont {Goriely}}, \bibinfo {author} {\bibfnamefont {F.}~\bibnamefont {Tondeur}},\ and\ \bibinfo {author} {\bibfnamefont {J.~M.}\ \bibnamefont {Pearson}},\ }\bibfield  {title} {\bibinfo {title} {{A Hartree-Fock Nuclear Mass Table}},\ }\href {https://doi.org/10.1006/adnd.2000.0857} {\bibfield  {journal} {\bibinfo  {journal} {Atom. Data Nucl. Data Tabl.}\ }\textbf {\bibinfo {volume} {77}},\ \bibinfo {pages} {311} (\bibinfo {year} {2001})}\BibitemShut {NoStop}%
\bibitem [{\citenamefont {Hilaire}\ \emph {et~al.}(2012)\citenamefont {Hilaire}, \citenamefont {Girod}, \citenamefont {Goriely},\ and\ \citenamefont {Koning}}]{Hilaire:2012tb}%
  \BibitemOpen
  \bibfield  {author} {\bibinfo {author} {\bibfnamefont {S.}~\bibnamefont {Hilaire}}, \bibinfo {author} {\bibfnamefont {M.}~\bibnamefont {Girod}}, \bibinfo {author} {\bibfnamefont {S.}~\bibnamefont {Goriely}},\ and\ \bibinfo {author} {\bibfnamefont {A.~J.}\ \bibnamefont {Koning}},\ }\bibfield  {title} {\bibinfo {title} {{Temperature-dependent combinatorial level densities with the D M-1 Gogny force}},\ }\href {https://doi.org/10.1103/PhysRevC.86.064317} {\bibfield  {journal} {\bibinfo  {journal} {Phys. Rev. C}\ }\textbf {\bibinfo {volume} {86}},\ \bibinfo {pages} {064317} (\bibinfo {year} {2012})}\BibitemShut {NoStop}%
\bibitem [{\citenamefont {Kopecky}\ and\ \citenamefont {Uhl}(1990)}]{Kopecky:1990zz}%
  \BibitemOpen
  \bibfield  {author} {\bibinfo {author} {\bibfnamefont {J.}~\bibnamefont {Kopecky}}\ and\ \bibinfo {author} {\bibfnamefont {M.}~\bibnamefont {Uhl}},\ }\bibfield  {title} {\bibinfo {title} {{Test of gamma-ray strength functions in nuclear reaction model calculations}},\ }\href {https://doi.org/10.1103/PhysRevC.41.1941} {\bibfield  {journal} {\bibinfo  {journal} {Phys. Rev. C}\ }\textbf {\bibinfo {volume} {41}},\ \bibinfo {pages} {1941} (\bibinfo {year} {1990})}\BibitemShut {NoStop}%
\bibitem [{\citenamefont {{Brink}}(1957)}]{1957NucPh...4..215B}%
  \BibitemOpen
  \bibfield  {author} {\bibinfo {author} {\bibfnamefont {D.~M.}\ \bibnamefont {{Brink}}},\ }\bibfield  {title} {\bibinfo {title} {{Individual particle and collective aspects of the nuclear photoeffect}},\ }\href {https://doi.org/10.1016/0029-5582(87)90021-6} {\bibfield  {journal} {\bibinfo  {journal} {Nucl. Phys.}\ }\textbf {\bibinfo {volume} {4}},\ \bibinfo {pages} {215} (\bibinfo {year} {1957})}\BibitemShut {NoStop}%
\bibitem [{\citenamefont {Axel}(1962)}]{Axel:1962zz}%
  \BibitemOpen
  \bibfield  {author} {\bibinfo {author} {\bibfnamefont {P.}~\bibnamefont {Axel}},\ }\bibfield  {title} {\bibinfo {title} {{Electric Dipole Ground-State Transition Width Strength Function and 7-Mev Photon Interactions}},\ }\href {https://doi.org/10.1103/PhysRev.126.671} {\bibfield  {journal} {\bibinfo  {journal} {Phys. Rev.}\ }\textbf {\bibinfo {volume} {126}},\ \bibinfo {pages} {671} (\bibinfo {year} {1962})}\BibitemShut {NoStop}%
\bibitem [{\citenamefont {Goriely}\ and\ \citenamefont {Khan}(2002)}]{Goriely:2002cx}%
  \BibitemOpen
  \bibfield  {author} {\bibinfo {author} {\bibfnamefont {S.}~\bibnamefont {Goriely}}\ and\ \bibinfo {author} {\bibfnamefont {E.}~\bibnamefont {Khan}},\ }\bibfield  {title} {\bibinfo {title} {{Large scale QRPA calculation of E1 strength and its impact on the neutron capture cross-section}},\ }\href {https://doi.org/10.1016/S0375-9474(02)00860-6} {\bibfield  {journal} {\bibinfo  {journal} {Nucl. Phys. A}\ }\textbf {\bibinfo {volume} {706}},\ \bibinfo {pages} {217} (\bibinfo {year} {2002})}\BibitemShut {NoStop}%
\bibitem [{\citenamefont {Goriely}\ \emph {et~al.}(2004)\citenamefont {Goriely}, \citenamefont {Khan},\ and\ \citenamefont {Samyn}}]{Goriely:2003rz}%
  \BibitemOpen
  \bibfield  {author} {\bibinfo {author} {\bibfnamefont {S.}~\bibnamefont {Goriely}}, \bibinfo {author} {\bibfnamefont {E.}~\bibnamefont {Khan}},\ and\ \bibinfo {author} {\bibfnamefont {M.}~\bibnamefont {Samyn}},\ }\bibfield  {title} {\bibinfo {title} {{Microscopic HFB + QRPA predictions of dipole strength for astrophysics applications}},\ }\href {https://doi.org/10.1016/j.nuclphysa.2004.04.105} {\bibfield  {journal} {\bibinfo  {journal} {Nucl. Phys. A}\ }\textbf {\bibinfo {volume} {739}},\ \bibinfo {pages} {331} (\bibinfo {year} {2004})}\BibitemShut {NoStop}%
\bibitem [{\citenamefont {Daoutidis}\ and\ \citenamefont {Goriely}(2012)}]{Daoutidis:2012iz}%
  \BibitemOpen
  \bibfield  {author} {\bibinfo {author} {\bibfnamefont {I.}~\bibnamefont {Daoutidis}}\ and\ \bibinfo {author} {\bibfnamefont {S.}~\bibnamefont {Goriely}},\ }\bibfield  {title} {\bibinfo {title} {{Large-scale Continuum Random Phase Approximation predictions of dipole strength for astrophysical applications}},\ }\href {https://doi.org/10.1103/PhysRevC.86.034328} {\bibfield  {journal} {\bibinfo  {journal} {Phys. Rev. C}\ }\textbf {\bibinfo {volume} {86}},\ \bibinfo {pages} {034328} (\bibinfo {year} {2012})}\BibitemShut {NoStop}%
\bibitem [{\citenamefont {Xu}\ \emph {et~al.}(2021)\citenamefont {Xu}, \citenamefont {Goriely},\ and\ \citenamefont {Khan}}]{Xu:2021xet}%
  \BibitemOpen
  \bibfield  {author} {\bibinfo {author} {\bibfnamefont {Y.}~\bibnamefont {Xu}}, \bibinfo {author} {\bibfnamefont {S.}~\bibnamefont {Goriely}},\ and\ \bibinfo {author} {\bibfnamefont {E.}~\bibnamefont {Khan}},\ }\bibfield  {title} {\bibinfo {title} {{Systematical studies of the E1 photon strength functions combining the Skyrme-Hartree-Fock-Bogoliubov plus quasiparticle random-phase approximation model and experimental giant dipole resonance properties}},\ }\href {https://doi.org/10.1103/PhysRevC.104.044301} {\bibfield  {journal} {\bibinfo  {journal} {Phys. Rev. C}\ }\textbf {\bibinfo {volume} {104}},\ \bibinfo {pages} {044301} (\bibinfo {year} {2021})}\BibitemShut {NoStop}%
\end{thebibliography}%

\end{document}